\documentclass{IEEEtaes}

\usepackage{color,array,amsthm}
\usepackage{graphicx}
\usepackage{amsmath}
\usepackage{amssymb}
\usepackage{amsfonts}
\usepackage{mathtools}
\usepackage{subcaption}
\usepackage{booktabs}
\usepackage{multirow}
\usepackage{algorithmic}
\usepackage{algorithm}

\jvol{61}
\jnum{6}
\jmonth{December}
\pubyear{2025}
\doiinfo{TAES.2025.3588823}

\newtheorem{remark}{Remark}
\usepackage{amsmath}
\usepackage{bm}
\usepackage{mathtools}
\usepackage{multirow}
\usepackage{multicol}
\usepackage{tabularx}
\usepackage{booktabs}

\newcommand{\func}[2]{#1(#2)}
\newcommand{\Func}[2]{#1\left(#2\right)}

\newcommand{\cond}[2]{{#1}|{#2}}

\newcommand{\linefrac}[2]{\left.{#1}\middle/{#2}\right.}

\newcommand{\normVec}[1]{\left\|#1\right\|}

\newcommand{\VecEng}[1]{\mathbf{#1}}
\newcommand{\MatEng}[1]{\mathbf{#1}}
\newcommand{\VecGrk}[1]{\bm{#1}}

\newcommand{\PD}[3][display]{
    \ifthenelse{\equal{#1}{line}}
        {\linefrac{\partial{#2}}{\partial{#3}}}
        {\frac{\partial{#2}}{\partial{#3}}}
}

\newcommand{\transpose}{\text{T}}

\newcommand{\hDB}{h_{{\text{DTED}}}}
\newcommand{\INS}{\text{INS}}
\newcommand{\true}{\text{true}}
\newcommand{\x}{\VecEng{x}}
\newcommand{\y}{\VecEng{y}}

\newcommand{\xk}{\x_{k}}
\newcommand{\yk}{\y_{k}}

\newcommand{\xkhor}{\x_{k}^{\text{hor}}}

\newcommand{\Gaussian}{\mathcal{N}}

\newcommand{\xki}{\x_{k}^{i}}
\newcommand{\mun}{\mu^{(n)}}
\newcommand{\sigman}{\sigma^{(n)}}
\newcommand{\pin}{\pi^{(n)}}
\begin{document}
\title{Contours-seeking Proposal Density Particle Filter and Resilient Terrain-referenced Navigation}

\author{Junwoo Park}
\affil{Korea Advanced Institute of Science and Technology, Daejeon, South Korea}
\author{Hyochoong Bang} 
\member{Member, IEEE}
\affil{Korea Advanced Institute of Science and Technology, Daejeon, South Korea}

\corresp{\itshape (Corresponding author: Hyochoong Bang).}

\authoraddress{Junwoo Park is now with Nearthlab, Inc., Seoul, South Korea (e-mail: \href{vividlibra@gmail.com}{vividlibra@gmail.com}). Hyochoong Bang is with the Aerospace Engineering Department, Korea Advanced Institute of Science and Technology, Daejeon, South Korea (e-mail: \href{hcbang@kaist.ac.kr}{hcbang@kaist.ac.kr}).}

\markboth{PARK ET AL.}{CONTOURS-SEEKING PROPOSAL DENSITY PF AND RESILIENT TRN}
\maketitle

\begingroup
\renewcommand\thefootnote{}
\footnotetext{\textcopyright~2025 IEEE. Personal use of this material is
permitted. Permission from IEEE must be obtained for all other uses, in any
current or future media, including reprinting/republishing this material for
advertising or promotional purposes, creating new collective works, for
resale or redistribution to servers or lists, or reuse of any copyrighted
component of this work in other works. 

This is the author's accepted manuscript of 
``Contours-Seeking Proposal Density Particle Filter and Resilient
Terrain-Referenced Navigation,''
\textit{IEEE Trans. Aerosp. Electron. Syst.},
vol.~61, no.~6, pp.~15627--15641, Dec.~2025,
doi: \href{https://doi.org/10.1109/TAES.2025.3588823}
{10.1109/TAES.2025.3588823}.}
\addtocounter{footnote}{-1}
\endgroup

\begin{abstract}
Auxiliary navigation systems are essential for the robust operation of aerial vehicles, particularly in self-contained frameworks like terrain-referenced navigation. However, challenges such as multimodal likelihoods, highly nonlinear terrain elevations, and unknown prediction biases result in highly multimodal and less predictable posterior distributions, leading to particle filter degeneration. This study addresses the numerical instability and degeneration of the particle filter approach by proposing a sampling strategy tailored to this problem. The approach introduces a Gaussian mixture random forcing mechanism, which nudges particles along terrain slopes and against biases towards the most probable terrain contours. Each mixture is associated with a mode of likelihood, enhancing adaptability to unmodeled terrain features. To further improve effectiveness, auxiliary sampling selectively applies this mixture sampling to probable particles, yielding a less degenerate and evenly weighted particle set. Numerical experiments demonstrate the effectiveness of the proposed method in reducing weight variance, improving effective sample size. In addition, the approach exhibits strong resilience under deteriorating scenarios, such as severe unknown prediction bias and multimodal measurement noise, ensuring long-term reliable particle filtering.
\end{abstract}


\section{Introduction}\label{sec:introduction}
G{\scshape lobal} navigation satellite systems (GNSS) mitigate drifts of inertial navigation systems (INS) with its bounded estimates, while INS compensates for sparse updates of GNSS with its short term robustness. Although GNSS is considered a standard fusion source for INS, its dependence upon remote signals often makes the integrated GNSS/INS vulnerable to signal reception issues, such as multipath, spoofing, jamming or ionospheric scintillation. Despite advances in robust signal processing \cite{Borio2017Robust, Vila-Valls2020Survey}, self-contained operation is difficult to secure when the GNSS is the sole aiding source.

Database-referenced navigation (DBRN) emerges as a robust alternative by leveraging self-contained sensors and onboard databases instead of external signals. Geophysical information such as gravity \cite{Lee2015Performance}, magnetic field \cite{Kim2019Approach}, terrain elevation \cite{Bergman1999Terrain, Musso2000Recent}, or visual imagery of the ground \cite{Koch2006Visionbased, Hong2021Particle, Park2024Visual}; can be rendered as databases for navigation. Sensors including accelerometers, magnetometers, radar altimeters (RA), and cameras enable DBRN systems to operate independently of external infrastructure.

\begin{figure}[!htbp]
    \centering
    \includegraphics[width=0.85\columnwidth]{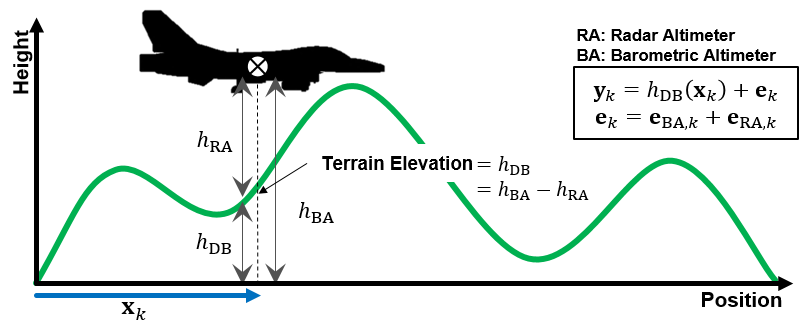}
    \caption{Principle of terrain-referenced navigation problem illustrated in 1-dimensional state space.}
    \label{fig:trn}
\end{figure}

Among DBRN approaches, terrain-referenced navigation (TRN) \cite{Park2017New}, also known as terrain-aided positioning (TAP) \cite{Nordlund2009Marginalized}, stands out for GNSS-denied or GNSS-challenging scenarios. Real-world implementations like terrain contour matching (TERCOM) \cite{Golden1980Terrain} and Sandia inertial terrain-aided navigation (SITAN) \cite{Hollowell1990Heli} demonstrate TRN's practicality. The TRN problem involves estimating a vehicle's position, $\xk$, using a series of terrain elevation measurements, $\y_{0:k}$. Under the Bayesian framework \cite{Bergman1999Terrain, Bergman1997Bayesian}, the task becomes estimating the posterior distribution $\func{p}{\cond{\xk}{\y_{0:k}}}$. Figure~\ref{fig:trn} illustrates the simplest form of the TRN problem with aids from a radar altimeter and a barometric altimeter, each responsible for terrain clearance and vehicle altitude.

\begin{figure}[!htbp]
    \centering
    \includegraphics[width=0.95\columnwidth]{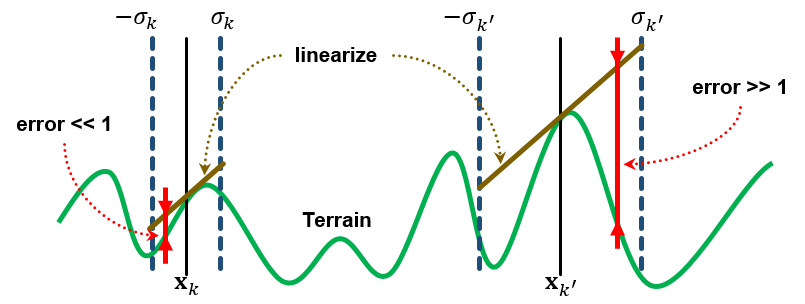}
    \caption{Nonlinear and ambiguous characteristics of terrain elevation captured within the estimated uncertainty.}
    \label{fig:nonlinear}
\end{figure}

Despite its simplicity, TRN presents significant challenges due to inherently nonlinear and ambiguous characteristics of terrain elevation. Traditional approaches like the extended Kalman filter (EKF) may not capture local terrain deviations within estimated error variance, leading to filter divergence. Figure~\ref{fig:nonlinear} highlights these issues.

Particle filters (PF) \cite{Gordon1993Novel} and point mass filters (PMF) \cite{Bergman1997Pointmassa} are commonly used for TRN due to their flexibility with nonlinearity and multimodality. While PMFs suffer from the curse of dimensionality, PFs utilize stochastic samples to approximate the state space, offering better scalability. This study focuses on PF-based approaches, which are more compatible with flexible INS integration due to their dimension-independent convergence properties. However, PFs still face challenges with particle degeneration, especially with multimodal measurement noise and biased odometry. In a particle filter, the posterior distribution over the state space $\mathcal{X}$ is approximated by a weighted sum of stochastic samples:

\begin{equation}
    \func{p}{\cond{\x}{\y}} \approx \sum_{i=1}^{N}{w^{i} \func{\delta}{\x - \x^{i}}},
\end{equation}

\noindent where $\y$ denotes the measurement, $\x^{i}$ and $w^{i}$ denote $i^\text{th}$ particle and its importance weight with the support of the Dirac-delta function, $\func{\delta}{\cdot}$. When the particle was sampled from an alternative proposal distribution, $\func{q}{\x}$, that is (preferably) easier to sample from, the weight is proportional to: 

\begin{equation} \label{eq:weight_property}
    w^{i}\propto\frac{\func{p}{\cond{\y}{\x^{i}}}\func{p}{\x^{i}}}{\func{q}{\x^{i}}},
\end{equation}

\noindent It is worth mentioning that ideally, if $\func{q}{\cdot}$ matches the true posterior, resultant weights play no role as all particles are equally valid.

Both theoretical analyses \cite{Doucet2000Sequential, VanLeeuwen2019Particle, Snyder2015Performance}; and empirical evidence showed that sequential multiplication of likelihoods increases the variance of importance weights over time, eventually causing filter collapse. The situations where the drawn particles are located at the thin tail of likelihood or where the particles do not fully occupy the high-likelihood regions are the typical journey to full degeneration. This becomes particularly severe with highly informative measurements or when the proposal distribution significantly deviates from the posterior.

\begin{figure}[!htbp]
    \centering
    \begin{subfigure}[b]{.3\textwidth}
        \includegraphics[width=\textwidth]{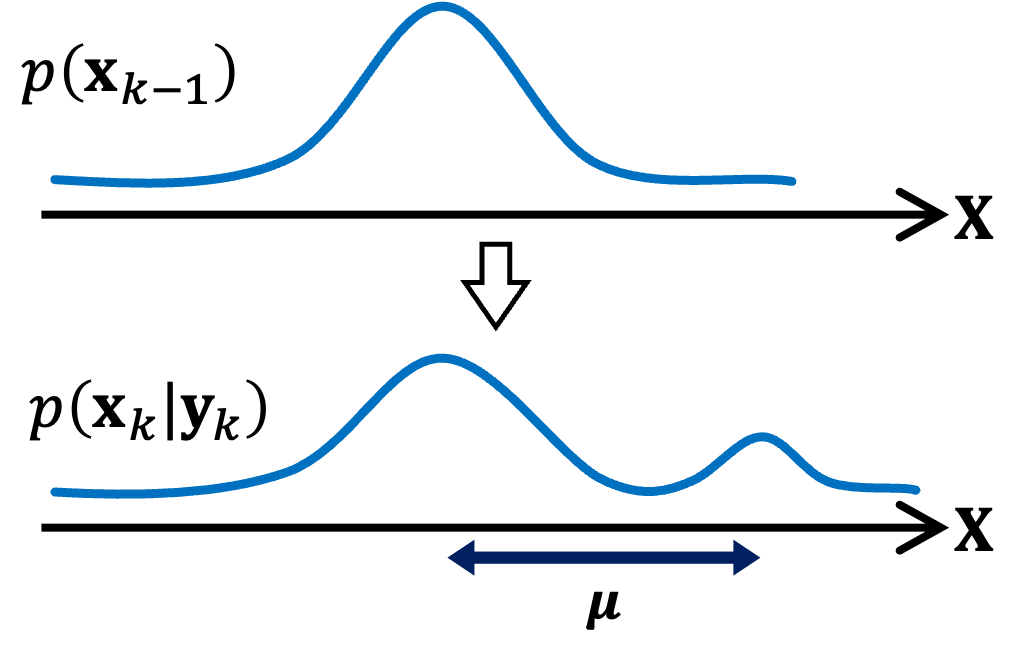}
        \caption{A linear measurement update example with $\yk=\xk+\VecEng{e}_{k}$.}
        \label{fig:problem_illust_1}
    \end{subfigure}
    \begin{subfigure}[b]{.3\textwidth}
        \includegraphics[width=\textwidth]{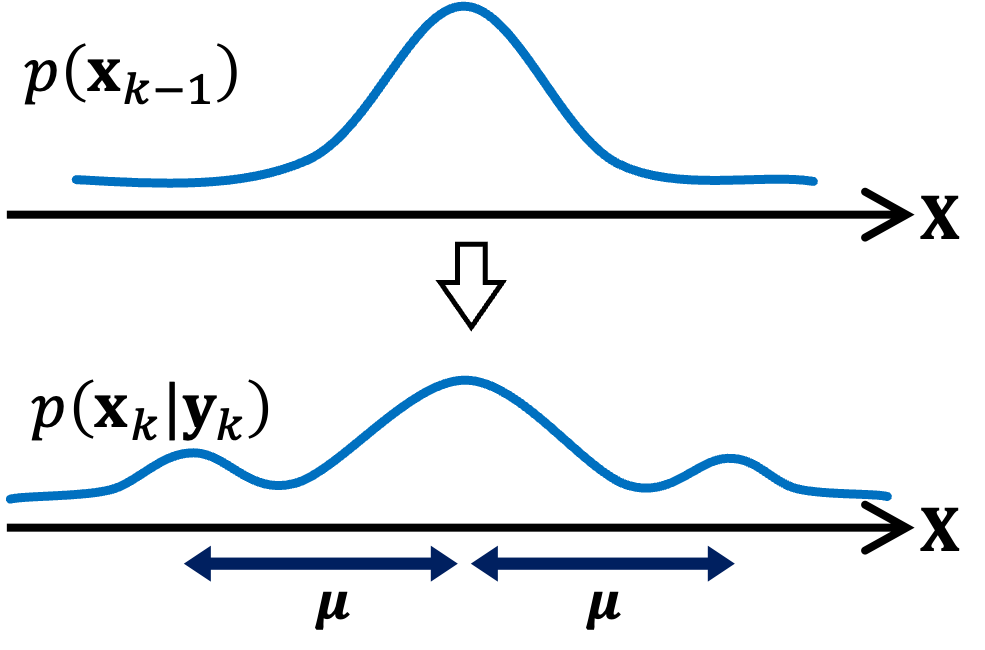}
        \caption{A nonlinear measurement update example with $\yk=|\xk|+\VecEng{e}_{k}$.}
        \label{fig:problem_illust_2}
    \end{subfigure}
    \begin{subfigure}[t]{.35\textwidth}
        \includegraphics[width=\textwidth]{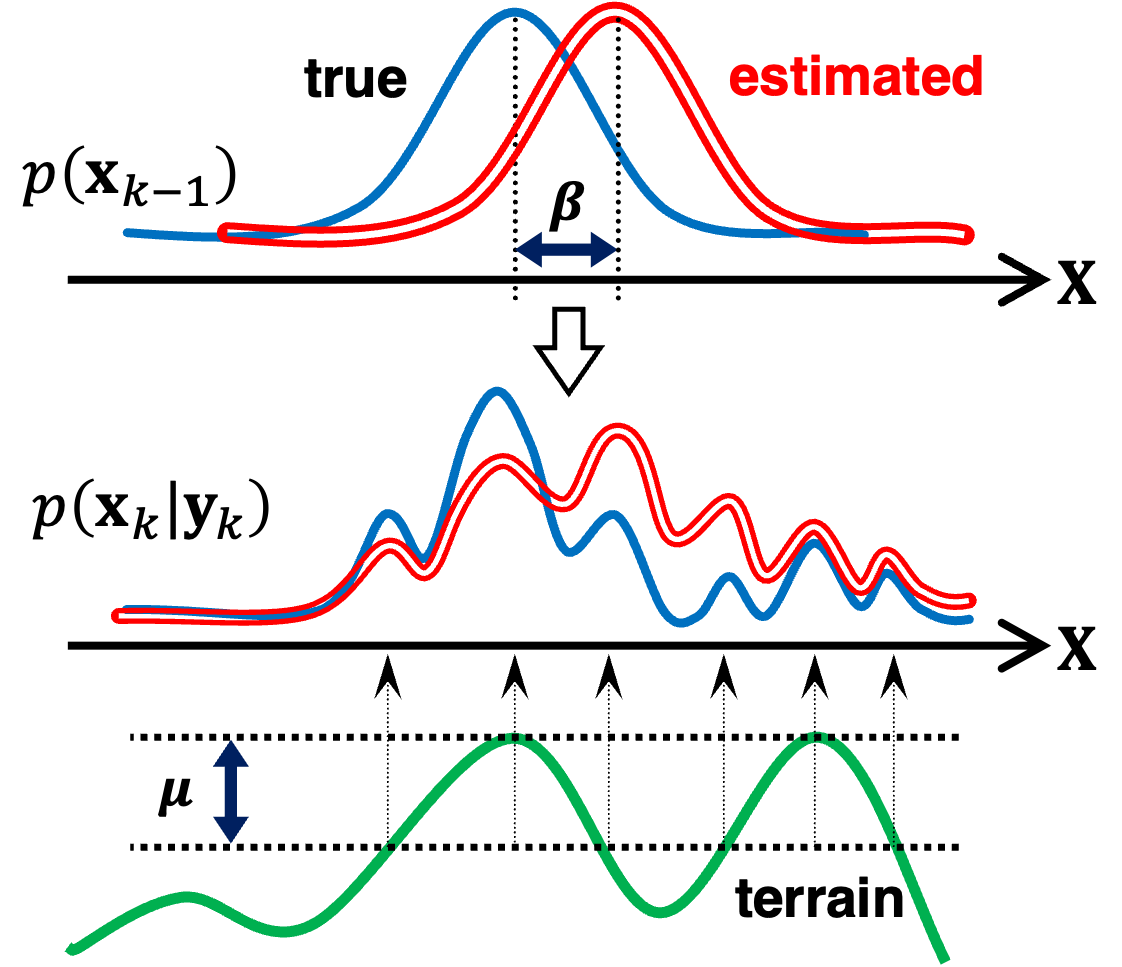}
        \caption{Highly nonlinear terrain measurement $\yk=\Func{\hDB}{\xk}+\VecEng{e}_{k}$ with prediction bias $\beta$.}
        \label{fig:problem_illust_3}
    \end{subfigure}
    \caption{Illustration of the effect of combined highly nonlinear measurement equation, multimodal likelihood, and prediction bias on the posterior distribution. The unknown prediction bias is denoted as $\beta$, and the measurement likelihood is bimodal Gaussian: one zero-mean and another centered at $\mu>0$.}
    \label{fig:problem_illust}
\end{figure}

Three characteristics of TRN exacerbate this vulnerability: radar altimeter measurements corrupted by multimodal noise \cite{Schon2005Marginalized}, highly nonlinear measurement equation, and INS's drift from deduced reckoning principle. As a combination of the former two, the true posterior exhibits pronounced multiple modes distributed across a broad area, making both the design of effective proposal distributions and the maintenance of filter stability extremely challenging. The incorrect prediction then deforms the estimated distribution in a less predictable way due to terrain's high nonlinearity, potentially leading the filter to degenerate. Figure~\ref{fig:problem_illust} demonstrates how these effects collectively distort the estimated posterior distribution from its desired form.

While intermittent resampling \cite{Li2015Resampling} can mitigate degeneracy, its stochastic nature offers limited theoretical guarantees. A more effective approach is preventing particle and weight degeneracy from the outset. Our approach is based on the two basic observations:
\begin{itemize}
    \item[O1] the modes of the posterior distribution align with \textit{terrain contours},
    \item[O2] PF's performance and stability gradually deteriorate without proper support from particles.
\end{itemize}
The O1 also aligns with the TERCOM's \cite{Golden1980Terrain} emphasis on contour matching. After all, the TRN problem is to find an inverse map from the sequential superposition of overlapping \textit{contours}, as described in \cite{Jang2017Acquisition}.

The strategy this study takes is straightforward: to consecutively locate particles around those contours so as not to largely deviate from the true posterior modes. 
We implement a local approach that nudges particles toward probable contours using terrain gradients taking deviated multimodality into account. The objectives of this study are accurate approximation of multimodal empirical filtering density and mitigation of sample degeneracy through maintenance of even-weight particles, by designing a proposal distribution with solid support for the posterior under the challenging scenario outlined previously. Thereby, the numerical instability of the PF in TRN problem is resolved or significantly delayed. The particle flow-based approach \cite{Li2017Particle} is ruled out due to the stiffness and multimodality of the TRN problem.

\subsection{Related Works And Contributions}\label{sec:related_works}
Various approaches have been proposed to address particle filter degeneracy in TRN applications. Merlinge et al. \cite{Merlinge2016Box, Merlinge2019Box}; introduced box-regularized particle filter (BRPF), which employed set membership formalism to represent particles as interval boxes rather than Dirac points. By maintaining pessimistic kernels in box form, BRPF accommodated a wider range of measurements across multiple particles. The approach achieved lower divergence rate even under the large initial error and effectively delayed degeneracy problem.

Teixeira et al. \cite{Teixeira2012Novel, Teixeira2017Robust} proposed an alternative approach combining prior and non-informative (uniform) distributions. However, the uniform distribution sampling suffered from determining interval size and exhibited poor scalability. Although a Fisher information-based criterion \cite{Ly2017Tutorial} was suggested, the approach was inefficient due to indiscriminate particle coverage expansion. The general regularization scheme \cite{Musso2001Improving} achieves similar objectives more systematically.

The mixture-regularized particle filter, introduced by Murangira et al. \cite{Murangira2011Robust, Murangira2016Mixture}, operated multiple PF instances, with each instance handling a distinct posterior mode to preserve the overall multimodal characteristics. The approach employed a set of maximum a posteriori (MAP)-based Gaussian proposal distributions assuming single modality of each PF instance via additional clustering methods. However, it required numerous heuristic parameters, such as clustering bandwidth, not revealed by the authors, and thus the results were hardly reproducible. Moreover, calculating MAP is computationally intensive and numerically instable at stiff problems, hence an error state design was necessitated. Meanwhile, The MAP is also leveraged in \cite{Musso2019Terrain} via Laplace PF.

These approaches revealed the importance of maintaining robust support for the posterior to minimize ineffective samples. Despite their original ideas, these methods have several limitations. They primarily addressed nonlinear and ambiguous terrain characteristics while overlooking the challenges posed by multimodal likelihood of RA and prediction biases. Furthermore, modifications to canonical particle filtering principles---whether through set membership \cite{Merlinge2016Box}, prior correction \cite{Teixeira2017Robust}, or regularization \cite{Musso2001Improving}---introduced additional tuning parameters that compromise practicality and scalability. Moreover, most existing work focused predominantly on estimation performance metrics. In contrast, our study examines both particle weight evenness and filter numerical stability to provide a more comprehensive evaluation framework.

Again, this study aims to develop a proposal distribution capable of handling the multimodality of RA and prediction biases compounded via highly nonlinear terrain. Based on \eqref{eq:weight_property}, a proposal distribution that closely approximates the true posterior produces more evenly weighted particles. The proposed method would thereby increase the number of effective samples, maximize and maintain them over extended periods. The key contributions of this study are:
\begin{enumerate}
    \item[C1] Development of a mode-associated mixture proposal accommodating the combination of RA's multimodality and highly nonlinear terrain.
    \item[C2] An exact Bayesian recursion-based approach where heuristics or tuning are unnecessary.
    \item[C3] Resilience against unknown time-varying INS biases, maintaining filter stability and accuracy through adaptive mechanism of particle sampling, without explicit bias estimation or compensation.
\end{enumerate}

These are achieved by a superposition of stochastic nudges guiding particles toward one of the probable terrain contours, that is further reinforced by the principle of the auxiliary variable sampling. We validate our approach through comprehensive analysis of weight variance, effective sample size, and weight entropy, demonstrating robust performance against typical causes of PF-TRN degeneration. Our parameter study advocates the exploitation of modality parameter for mode (contour)-associated random forcing. Additionally, we demonstrate convergence under challenging scenarios involving time-varying unknown biases, establishing the method's practical viability for real-world TRN applications.

\subsection{Outline}\label{sec:outline}
Section~\ref{sec:TRNandDTED} presents the mathematical foundations of TRN focusing on multimodal characteristics of RA arising from unmodeled terrain. Section~\ref{sec:PFandQ} reviews the principle of PF and several related topics, especially in terms of the proposal distribution and optimality. Section~\ref{sec:EWPFTRN} highlights the proposed approach addressing practical PF-TRN challenges. Section~\ref{sec:experiment} presents extensive numerical validation under diverse conditions, parameter studies and useful derived features of the proposed approach. Section~\ref{sec:conclusion} summarizes key findings and implications.

\section{Terrain-referenced Navigation and Digital Terrain Elevation Database}\label{sec:TRNandDTED}
\subsection{System Model}\label{sec:system_model}
The TRN problem is posed to estimate the state variable:
\begin{equation} \label{eq:state}
    \x_{k} = \begin{bmatrix} L_{k} & \lambda_{k} & h_{k} \end{bmatrix}^{\transpose},
\end{equation}

\noindent where $L$, $\lambda$, and $h$ represent latitude, longitude and altitude, respectively. While a full INS mechanization model can be exploited \cite{Nordlund2009Marginalized}, we focus on the simplified position kinematics augmented with unknown bias term to capture drifting characteristics of INS. The simplified INS-aided incremental can be modeled as \cite{Bergman1999Terrain, Park2022Evenly} 

\begin{equation} \label{eq:propagation_model}
    \begin{aligned}
        \xk &= \func{f}{\x_{k-1}} + \VecEng{v}_k \\
            &= \x_{k-1} + \Delta\xk^{\INS} + \VecEng{v}_k,
    \end{aligned}
\end{equation}

\noindent where $\VecEng{v}_k$ represents additive zero-mean Gaussian noise with $\mathbb{E}[\VecEng{v}_{k}\VecEng{v}_{k'}^{\transpose}]= \func{\delta}{k-k'}\MatEng{Q}$, and $\Delta\xk^{\INS}$ denotes the INS position increment:

\begin{equation} \label{eq:ins_model}
    \Delta\xk^{\INS} = \xk^{\true} - \x_{k-1}^{\true} + \VecEng{b}_{k}.
\end{equation}

\noindent The bias $\VecEng{b}_k$ is uninformed to the filter. An uncorrected estimator based on \eqref{eq:ins_model} will diverge following the characteristics of $\VecEng{b}_{k}$; even a corrected PF would continuously shift from the true distribution in the sequel. 

For concise and regularized numerical validation of proposed approach under various conditions, we model $\VecEng{b}_k$ as unknown and time-varying as

\begin{equation}\label{eq:bias_model}
    \VecEng{b}_{k} = \VecEng{b}_{v} + (k-\func{s}{k})\VecEng{b}_{a}\func{\mathbb{U}_{\func{s}{k}}}{k},
\end{equation}

\noindent with fixed unknown biases of velocity $\VecEng{b}_{v}$ and acceleration $\VecEng{b}_{a}$. Here, $\func{\mathbb{U}_{\tau}}{x}$ denotes a step function, i.e., 1 when $x\geq\tau$ and 0 otherwise, representing unknown fault of IMU sensors \cite{Avram2017Quadrotor}. $\func{s}{k}$ represents unknown schedules, where any finite constant value $k_1$ during an interval $k\in[k_1,k_2], k_1<k_2$ will induce linear increments. Note that the position deviation is linear with respect to $\VecEng{b}_{v}$ and quadratic with respect to the second term.

\begin{remark}
    This study focuses on how sampling affects the position estimates of PF TRN. The simplified model \eqref{eq:propagation_model}, \eqref{eq:ins_model}, and \eqref{eq:bias_model} is considered passive assuming the positional increment is simply provided, and neither velocities nor biases are estimated. High fidelity INS model and tight fusion with TRN can be found in \cite{Nordlund2009Marginalized}.
\end{remark}

The vertical clearance becomes TRN's measurement and is modeled as

\begin{equation} \label{eq:measurement_model}
    \begin{aligned}
        \yk &= \func{h}{\x_{k}} + \VecEng{e}_{k} \\
            &= h_{k} - \func{\hDB}{L_{k}, \lambda_{k}} + \VecEng{e}_{k},
    \end{aligned}
\end{equation}

\noindent where $\VecEng{e}_k$ exhibits collective error characteristics of RA, digital terrain elevation data (DTED), and barometer measurements if present. $\hDB$ denotes terrain database lookup operations.

Even if we assume Gaussianity on $\VecEng{e}_k$, i.e., $\mathbb{E}[\VecEng{e}_{k}\VecEng{e}_{k'}^{\transpose}]= \func{\delta}{k-k'}\MatEng{R}$, the TRN inherently exhibits multimodality due to terrain ambiguity. This characteristic renders Gaussian estimators unsuitable. Moreover, multimodality of RA measurement highlighted in the following section makes the numerically stable estimation more challenging. We address these challenges, that is even worsened in conjunction with unknown $\VecEng{b}_{k}$, in Section~\ref{sec:EWPFTRN}.

\subsection{Multimodality of Radar Altimeter}\label{sec:RA_mm}
\begin{figure*}[!htbp]
    \centering
    \includegraphics[width=0.85\textwidth]{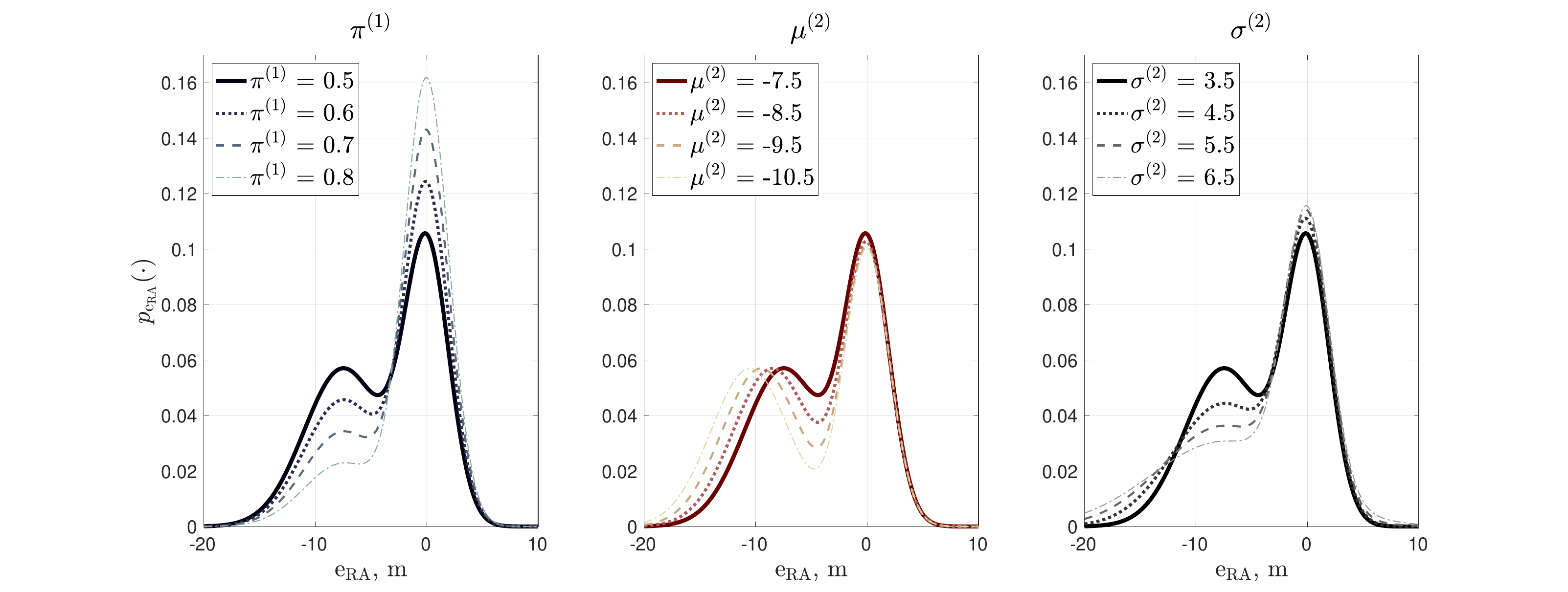}
    \caption{Multimodality of RA measurement noise introduced by unmodeled terrain.}
    \label{fig:ra_gmm_noise}
\end{figure*}

RA used in TRN applications typically operate in the C-band frequency range \cite{Campbell2006Application}. Since their wavelengths are on the order of a few centimeters, these signals rarely penetrate vegetation, instead scattering from canopies and treetops \cite{Park2020Parameter}. Consequently, terrain clearance reading from RA and terrain elevation at the estimated position do not sum to the vehicle altitude. Therefore stochastically, an additional mode is developed in RA measurement noise. This phenomenon was also observed in field experiments and documented in \cite{Schon2005Marginalized}. The following Gaussian mixture model (GMM) should better describe the multimodal characteristics of $\VecEng{e}_{k}$:

\begin{equation} \label{eq:ra_noise}
    \VecEng{e}_{\text{RA}, k}
        \sim\sum_{n=1}^{N_m}{\pi^{(n)}\Func{\mathcal{N}}{\mun_{k}, \left(\sigman_{k}\right)^{2}}},
\end{equation}

\noindent where $N_{m}$ denotes the number of modes, which is typically 2. $\mu_k^{(1)}$ equals zero and $\mu_k^{(2)}$ is a distinctive negative value corresponding to the average (negative) height of local natural obstacles. The parameters $\sigma_{k}^{(n)}$ denote the standard deviations of respective modes. This model also effectively captures real-world scenarios where the DTED was built in the past, and the time lag from which to the operation time is significant. Figure~\ref{fig:ra_gmm_noise} illustrates RA noise multimodality under varying parameters of \eqref{eq:ra_noise}. Three solid lines use the identical parameters: $\pi^{(1)}=0.5, \mu^{(2)}=-7.5, \sigma^{(1)}=2, \sigma^{(1)}=3.5$.

These latent parameters depend on terrain classification, e.g., bare land or dense forest. Factors such as leaf area index (LAI) \cite{Rosen1996Surface}, weather conditions, and humidity influence these parameters \cite{Park2020Parameter, Prevot1993Estimating}. This study assumes prior knowledge of these values with GMM parameter estimation, as addressed in \cite{Park2020Parameter}. Then, deriving a proposal density that can preserve effective samples by fully exploiting the model knowledge is addressed.

Note that when barometer aiding is assumed and the altitude is stabilized, the problem is narrowed down to estimating $[L_{k}, \lambda_{k}]^{\transpose}$ with a slightly modified measurement model: $\yk = \func{\hDB}{L_{k}, \lambda_{k}} + \VecEng{e}_{k}$. Then, $\VecEng{e}_{k}$ exhibits negated modes and incorporates barometer noise $\sigma_{\text{BA},k}$:

\begin{equation} \label{eq:measurement_noise}
    \func{p_{\VecEng{e}_{k}}}{\cdot}
        =\sum_{n=1}^{N_m}{\pi^{(n)}\Func{\mathcal{N}}{\cdot~;-\mun_{k}, \left(\sigman_{k}\right)^{2}+\sigma_{\text{BA},k}^{2}}}.
\end{equation}

\noindent The positive modes then represent the local canopy's average height and variance.

\subsection{Digital Terrain Elevation Database Interpolation}\label{sec:dted}
Let $L_{ij}, \lambda_{ij}$ for $i,j\in\{0, 1\}$ denote four corner points that envelop the query point using the floor operator $\lfloor\cdot\rfloor$ as

\begin{equation} \label{eq:cell}
    \begin{bmatrix}
        L_{00}\\
        L_{11}\\
        \lambda_{00} \\
        \lambda_{11} \\
    \end{bmatrix} \coloneqq
    \begin{bmatrix}
        \func{L}{\left\lfloor{(\func{L}{0} - L_{k})}/{\Delta L}\right\rfloor} \\
        \func{L}{\left\lfloor{(\func{L}{0} - L_{k})}/{\Delta L}\right\rfloor + 1}\\
        \func{\lambda}{\left\lfloor{(\lambda_{k} - \func{\lambda}{0})}/{\Delta\lambda}\right\rfloor} \\
        \func{\lambda}{\left\lfloor{(\lambda_{k} - \func{\lambda}{0})}/{\Delta\lambda}\right\rfloor+1}
    \end{bmatrix},
\end{equation}

\noindent where $\Delta$ values are resolutions of DTED, and $L(j)$ (or $\lambda(j)$) represent the $j$-th grid element. 

Based on the principle of bilinear interpolation, terrain elevation at arbitrary position is approximated as
\begin{equation} \label{eq:interpolation}
    \func{\hDB}{L_{k}, \lambda_{k}} \coloneqq h_{xy}xy + h_{x}x + h_{y}y + h_{00},
\end{equation}
\noindent where $x, y\in[0,1]$ are non-dimensional scalars parameterizing $L_{k}, \lambda_{k}$:
\begin{equation} \label{eq:parametrize}
    \begin{aligned}
        L_{k} &= L_{00} - \Delta L x, \\
        \lambda_{k} &= \lambda_{00} + \Delta\lambda y.
    \end{aligned}
\end{equation} 
\noindent The sign convention is intentional due to the indexing mismatch between matrix-like data structure and latitude. 

The coefficients of \eqref{eq:interpolation} are defined as 
\begin{equation} \label{eq:dted_coefficients}
    \begin{aligned}
        h_{xy}&\coloneqq h_{00}-h_{10}-h_{01}+h_{11},\\
        h_{x}&\coloneqq h_{10}-h_{00},  h_{y}\coloneqq h_{01}-h_{00},
    \end{aligned}
\end{equation}
\noindent given $h_{ij}=\func{\hDB}{L_{ii}, \lambda_{jj}}$. While advanced techniques such as cubic-spline or inverse distance weighting (IDW) \cite{Setianto2015Comparison} offer smoother elevation mapping, \eqref{eq:interpolation} suffices for the purpose of this study that is to develop a proposal distribution which pushes or pull particles towards probable contours according to the local terrain slope. It also provides fair computational efficiency.
\begin{remark}
    For a given fixed elevation, \eqref{eq:interpolation} represents a \textit{contour} analogous to a parabolic curve \cite{Jang2017Acquisition}. The 2D TRN is then to find an inverse map from time-varying parabolic curves to overlapping positions. 
\end{remark}

\section{Particle Filtering and Proposal Distribution}\label{sec:PFandQ}
\subsection{Principle of Particle Filter}\label{sec:PFBasics}
When it comes to the recursive formalism of the PF, particles and their weights approximate the filtering density:

\begin{equation}
    \func{p}{\cond{\xk}{\y_{0:k}}}\approx\sum_{i=1}^{N}{w_{k}^{i}\func{\delta}{\xk-\xki}},
\end{equation}

\noindent where $w_k^{i}$ represents the importance weight when sampling from a distribution other than the true posterior. The expectation of any function of interest $\func{g}{\cdot}$ over the posterior, $\func{p}{\cond{\xk}{\y_{0:k}}}$, can be approximated using Monte Carlo integration:

\begin{equation}
    \int_{\mathcal{X}_k}{\func{g}{\xk}\func{p}{\cond{\xk}{\y_{0:k}}}d\xk} \approx 
        \sum_{i=1}^{N}{w_{k}^{i} \func{g}{\xki}},
\end{equation}

\noindent with the minimum mean square error (MMSE) estimate as a notable example:

\begin{equation} \label{eq:mmse}
    \xk^{\text{MMSE}} = \int_{\mathcal{X}_k}{\xk\func{p}{\cond{\xk}{\y_{0:k}}}d\xk} \approx 
        \sum_{i=1}^{N}{w_{k}^{i} \xki}.
\end{equation}

The Bayesian recursion based on the Chapman-Kolmogorov equations,

\begin{equation} \label{eq:CK}
    \begin{split}
        \func{p}{\cond{\xk}{\y_{0:k-1}}} &= \int_{\mathcal{X}_{k-1}}{\func{p}{\cond{\xk}{\x_{k-1}}}\func{p}{\cond{\x_{k-1}}{\y_{0:k-1}}}d\x_{k-1}},\\
        \func{p}{\cond{\xk}{\y_{0:k}}} &= \frac{\func{p}{\cond{\yk}{\xk}}\func{p}{\cond{\xk}{\y_{0:k-1}}}}{\int_{\mathcal{X}_k}{\func{p}{\cond{\yk}{\xk}}\func{p}{\cond{\xk}{\y_{0:k-1}}}d\xk}},
    \end{split}
\end{equation}

\noindent becomes tractable by transforming the state space integration into a weighted summation of discrete elements.

Since direct sampling from the exact posterior is generally impossible \cite{Doucet2000Sequential}, consider now drawing particles from a proposal distribution of a general form, $\func{q}{\cond{\xk}{\x_{0:k-1}, \y_{0:k}}}$, sequentially, so that its temporal stack amounts to the full posterior:

\begin{equation}
    \Func{q}{\cond{\x_{0:k}}{\y_{0:k}}}=\Func{q}{\cond{\x_{0}}{\y_{0}}}\prod_{t=1}^{k}{\func{q}{\cond{\xk}{\x_{0:t-1}, \y_{0:t}}}}.
\end{equation}

\noindent Then, for each measurement $\yk$ and drawn sample $\xki\sim\func{q}{\cdot}$, the weight recursion follows:

\begin{equation} \label{eq:weight_update_general}
    w_k^{i} \propto w_{k-1}^{i}\frac{\Func{p}{\cond{\yk}{\xki}}\Func{p}{\cond{\xki}{\x_{k-1}^{i}}}}{\Func{q}{\cond{\xki}{\x_{0:k-1}^{i}, \y_{0:k}}}}.
\end{equation}

\noindent As noted in \eqref{eq:weight_property}, sampling directly from the filtering distribution would result in equally valid weights given $w_{0}^{1:N}=1/N$.

The prediction step of the PF is to occupy the state space by drawing samples $\{\xki\}_{i=1}^{N}$ according to some proposal distributions of our choosing and compensate for the mismatch using the importance weight for the drawn sample, $\xki$, as

\begin{equation} \label{eq:weight_prediction}
    w_{k|k-1}^{i} = w_{k-1}^{i} \frac{\Func{p}{\cond{\xki}{\x_{k-1}^{i}}}}{\Func{q}{\cond{\xki}{\x_{0:k-1}^{i}, \y_{0:k}}}}, 
\end{equation}

\noindent followed by the measurement update according to the Bayes' rule:

\begin{equation} \label{eq:weight_update_normalize}
    w_{k}^{i} = \frac{w_{k|k-1}^{i}\Func{p}{\cond{\yk}{\xki}}}{\sum_{j=1}^{N}{w_{k|k-1}^{j}\Func{p}{\cond{\yk}{\xk^{j}}}}},
\end{equation}

\noindent whose combination with \eqref{eq:weight_prediction} is analogous to \eqref{eq:weight_property} or \eqref{eq:weight_update_general}. 

A common choice for $q$ is the prior, $\func{q}{\cond{\xk}{\x_{0:k-1}^{i}, \y_{0:k}}}= \func{p}{\xk|\x_{k-1}^{i}} = \func{\mathcal{N}}{\func{f}{\x_{k-1}^{i}}, \MatEng{Q}_k}$ that nullifies \eqref{eq:weight_prediction}. Since weight recursion directly depends on drawn samples, certain proposal distributions may be more advantageous for the property of derived weights. This study focuses on tailoring $q$ for the TRN problem described previously.

\subsection{Optimal Proposal Distribution using Locally Linearized Measurement Model}\label{sec:optimal_q}
Research has established that importance weight variance inevitably increases over time \cite{Doucet2000Sequential}. This creates a fundamental trade-off between proposal weight \eqref{eq:weight_prediction} and likelihood weight \eqref{eq:weight_update_normalize}; a proposal distribution in strong favor of only the prior would gain much proposal weight while losing likelihood weight and vice versa, yielding high-variance of weights. Given that the signal-to-noise ratio (SNR) is approximately proportional to $\linefrac{\normVec{\MatEng{Q}}}{\normVec{\MatEng{R}}}$ \cite{Gustafsson2010Particle}, there exists an optimal proposal distribution that minimizes weight variance: a balance similar to what the EKF achieves in linear Gaussian systems.

Previous researches \cite{Doucet2000Sequential, Snyder2015Performance} have shown that the optimal proposal distribution in the PF framework, minimizing weight variance, takes the form:

\begin{equation} \label{eq:optimal_q}
    \func{q^*}{\cond{\xk}{\x_{k-1}^{i}, \yk}} = \func{p}{\cond{\xk}{\x_{k-1}^{i}, \yk}},
\end{equation}

\noindent which exhibits the convenient property:

\begin{equation} \label{eq:optimal_q_property}
    \begin{aligned}
        \func{p}{\cond{\xk}{\x_{k-1}^{i}, \yk}}
            &=\frac{\func{p}{\cond{\yk}{\xk,\x_{k-1}^{i}}}\func{p}{\cond{\xk}{\x_{k-1}^{i}}}}{\Func{p}{\cond{\yk}{\x_{k-1}^{i}}}} \\
            &=\frac{\func{p}{\cond{\yk}{\xk}}\func{p}{\cond{\xk}{\x_{k-1}^{i}}}}{\Func{p}{\cond{\yk}{\x_{k-1}^{i}}}}.
    \end{aligned}
\end{equation}

Direct substitution into \eqref{eq:weight_update_general} yields $w_{k}^{i}\propto w_{k-1}^{i}\func{p}{\cond{\yk}{\x_{k-1}^{i}}}$, indicating that weight variance becomes independent of sampling. Nevertheless, both the evaluation of $\func{p}{\cond{\yk}{\x_{k-1}^{i}}}=\int_{\mathcal{X}_{k}}{\func{p}{\cond{\yk}{\xk}}\func{p}{\cond{\xk}{\x_{k-1}^{i}}}d\xk}$ and sampling from $q^*$ are generally intractable for nonlinear non-Gaussian problems \cite{Doucet2001Sequential, Snyder2015Performance}. The key idea of incorporating the latest available information when sampling particles remains valid.

Following \cite{VanLeeuwen2019Particle}, we consider steering particles via:

\begin{equation} \label{eq:f_relaxation}
    \xk = \func{f}{\x_{k-1}} + \func{\MatEng{K}_k}{\yk - \Func{h}{\Func{f}{\x_{k-1}}}} + \hat{\VecEng{v}}_{k},
\end{equation}

\noindent where $\hat{\VecEng{v}}_{k}$ is an additive noise term (distinct from $\VecEng{v}_{k}$), and $\MatEng{K}$ denotes a gain multiplied to the innovation to nudge each particle toward the given measurement. This approximation of \eqref{eq:optimal_q} reduces to determining appropriate gain and noise distributions. Once we constrain the candidates for the proposal distribution into parametric a distribution that is easy to sample from, a simple Gaussian model should be favored:

\begin{equation} \label{eq:random_forcing}
    \xki\sim
        \Func{\mathcal{N}}{\Func{f}{\x_{k-1}^{i}} + \MatEng{K}_{k}^{i}\left(\yk - \hat{\y}_{k}^{i}\right), \left(\MatEng{I} - \MatEng{K}_{k}^{i}\MatEng{H}_{k}^{i}\right)\MatEng{Q}_{k-1}},
\end{equation}

\noindent with $\hat{\y}_{k}^{i} = \func{h}{\func{f}{\x_{k-1}^{i}}}$ and locally linearized measurement model:

\begin{equation} \label{eq:Jacobian}
    \MatEng{H}_{k}^{i} = \left.\PD{\func{h}{\xk}}{\xk}\right|_{\xk=\func{f}{\x_{k-1}^{i}}}.
\end{equation}

This formulation derives from standard EKF equations:

\begin{equation} \label{eq:ekf}
    \begin{aligned}
        \hat{\x}_{k-1|k-1}^{i} &\coloneqq \x_{k-1}^{i}, \\
        \hat{\x}_{k|k-1}^{i} &\sim \Func{\mathcal{N}}{\func{f}{\hat{\x}_{k-1|k-1}^{i}}, \MatEng{P}_{k|k-1}}, \MatEng{P}_{k|k-1} = \MatEng{Q}_{k-1}, \\
        \MatEng{K}_{k}^{i} &= \MatEng{P}_{k|k-1}\MatEng{H}_{k}^{i,\transpose} \left(\MatEng{H}_{k}^{i}\MatEng{P}_{k|k-1} \MatEng{H}_{k}^{i,\transpose} + \MatEng{R}_{k}\right)^{-1},
    \end{aligned}
\end{equation}

\noindent assuming uni-Gaussian measurement noise $\VecEng{e}_{k}\sim\Func{\mathcal{N}}{\VecEng{0},\MatEng{R}_{k}}$. This approach drives particles toward measurements via innovation, while the amount of \textit{random forcing} is determined by Kalman correction, treating each Dirac point (particle) of the prior as a zero-variance Gaussian.

Note that what the EKF assists in this context is the amount of random forcing, whereas the entire estimation is still carried out via particle filtering. This approximation is indeed the optimal proposal distribution \cite{VanLeeuwen2019Particle}, i.e., least varianced, for the linear Gaussian problem. However, the measurement equation of TRN is severely nonlinear with non-Gaussian additive noise as given in \eqref{eq:measurement_noise}. This study proposes a remedy for it in Section \ref{sec:auxgmm}.

\subsection{Particle Filtering using Auxiliary Variable}\label{sec:APF}
The auxiliary particle filter (APF) \cite{Pitt1999Filtering} offers an alternative approximation to $q^*$ by introducing sampling from a joint density $\func{q}{\xk, j|\y_{0:k}}$. Rather than directly sampling particles $\{\xki\}_{i=1}^{N}$, APF samples pairs $\{\xki, j^i\}_{i=1}^{N}$, where $j^i$ represents the index of the previous timestep particle from which $\xki$ is generated. This auxiliary index $j^i$ is discarded after sampling. The mathematical foundation of APF can be derived from Bayes' rule \cite{Arulampalam2002Tutorial}:

\begin{equation} \label{eq:APF_feature}
    \begin{aligned}
        \func{p}{\cond{\xk, j}{\y_{0:k}}} 
            & \propto \func{p}{\cond{\yk}{\xk}}\func{p}{\cond{\xk, j}{\y_{0:k-1}}} \\
            & = \func{p}{\cond{\yk}{\xk}}\func{p}{\cond{\xk}{j, \y_{0:k-1}}}\Func{p}{\cond{j}{\y_{0:k-1}}} \\
            & = \func{p}{\cond{\yk}{\xk}}\func{p}{\cond{\xk}{\x_{k-1}^{j}}}w_{k-1}^{j},
    \end{aligned}
\end{equation}

\noindent where discarding $j$ from the pair $\left(\xk,j\right)$ yields the desired particle set $\{\xki\}_{i=1}^{N}$. APF implements a proposal density satisfying:

\begin{equation}\label{eq:APF_proposal}
    \func{q}{\cond{\xk, j}{\y_{0:k}}} \propto \func{p}{\cond{\yk}{\VecGrk{\varphi}_{k}^{j}}}\func{p}{\cond{\xk}{\x_{k-1}^{j}}}w_{k-1}^{j},
\end{equation}

\noindent where $\VecGrk{\varphi}_{k}^{j}$ characterizes $\func{p}{\xk}$ given $\x_{k-1}^{j}$, typically through either the mean $\VecGrk{\varphi}_{k}^{i}=\mathbb{E}\left[\xk|\x_{k-1}^{i}\right]$ or a sample, i.e., $\VecGrk{\varphi}_{k}^{i}\sim\func{p}{\xk|\x_{k-1}^{i}}$. Given this, the weight recursion shown in \eqref{eq:weight_update_general} simplifies to:

\begin{equation}\label{eq:weight_update_apf}
    w_{k}^{i} 
        \propto w_{k-1}^{j^i} \frac{\func{p}{\cond{\yk}{\xki}}\func{p}{\cond{\xki}{\x_{k-1}^{j^i}}}}{\func{q}{\cond{\xki, j^i}{\y_{0:k}}}} 
        = \frac{\func{p}{\cond{\yk}{\xki}}}{\func{p}{\cond{\yk}{\VecGrk{\varphi}_{k}^{j^i}}}},
\end{equation}

\noindent with many terms canceling out.

The APF sampling process can be understood as generating $\{\xki\}_{i=1}^{N}$ from $\{\x_{k-1}^{i}\}_{i=1}^{N}$ by taking the relative importance of $j$-th particle in $(k-1)$-th timestep proportional to the right-hand side of \eqref{eq:APF_proposal}. Specifically, indices are drawn from a categorical distribution with probability mass function:

\begin{equation}\label{eq:categorical_distribution}
    \func{p}{\mathcal{J}=j} = \func{p}{\cond{\yk}{\VecGrk{\varphi}_{k}^{j}}}w_{k-1}^{j},
\end{equation}

\noindent and draws a particle from the distribution $\func{p}{\cond{\xk}{\x_{k-1}^{j}}}$ of the designated index. Basically, the APF simulates only the most probable particles in terms of generating $\yk$ by sampling from the entire mixture given in \eqref{eq:APF_proposal}.

\section{Even-Weights Particle Filter for Terrain-referenced Navigation}\label{sec:EWPFTRN}
The proposed approach of this study is straightforward: to associate a stochastic forcing given in Section~\ref{sec:optimal_q} with each mode of \eqref{eq:measurement_noise} and to apply the association to the likely particles based on the principle of the auxiliary sampling given in Section~\ref{sec:PFandQ}.\ref{sec:APF}. 

\subsection{Linearization of Terrain Elevation}\label{sec:linearize}
Before developing the mode-associated version of \eqref{eq:random_forcing}, we must address the calculation of the Jacobian for the Kalman correction. Let $\VecGrk{\theta}=[x, y]^{\transpose}$ represent the parametrization variables for $\xkhor = [L_{k}, \lambda_{k}]^{\transpose}$. Applying the chain rule to \eqref{eq:interpolation} yields:

\begin{equation}\label{eq:linearization_analytic}
    \begin{aligned}
        \MatEng{H}_{k}^{\text{hor}} &= \nabla_{\VecGrk{\theta}}h\nabla_{\xkhor}\VecGrk{\theta} \\
        &= \begin{bmatrix}
            h_{xy}y + h_{x} & h_{xy}x + h_{y}
        \end{bmatrix}
        \begin{bmatrix}
            -\Delta{L} & 0 \\
            0 & \Delta\lambda
        \end{bmatrix}^{-1} \\
        &= \left(h_{xy}\left(\xkhor\right)^{\transpose}\tilde{\Delta}^{-1} +
        \begin{bmatrix}
            h_{x}-\frac{h_{xy}\lambda_{00}}{\Delta\lambda} \\ h_{y}+\frac{h_{xy}L_{00}}{\Delta{L}}
        \end{bmatrix}^{\transpose}\right)
        \Delta^{-1},
    \end{aligned}
\end{equation}

\noindent where $\Delta=\Func{\text{diag}}{-\Delta{L}, \Delta\lambda}$ and $\tilde{\Delta}$ is its anti-diagonal version ($\tilde{\Delta}_{1,2}=-\Delta{L}, \tilde{\Delta}_{2,1}=\Delta\lambda$). The second equality follows from the inverse relationship of \eqref{eq:parametrize}. The third equation emphasizes that $\xk$ is the only ingredient in calculating the Jacobian and a single DTED lookup is required. Substituting this into \eqref{eq:Jacobian} yields:

\begin{equation} \label{eq:Jacobian_analytic}
    \MatEng{H}_{k}^{i, \text{hor}} =
        \left(h_{xy}{\func{f}{\x_{k-1}^{i}}}^{\transpose, \text{hor}}\tilde{\Delta}^{-1} + 
        \begin{bmatrix}
            h_{x}-\frac{h_{xy}\lambda_{00}}{\Delta\lambda} \\ h_{y}+\frac{h_{xy}L_{00}}{\Delta{L}}
        \end{bmatrix}^{\transpose}\right)
        \Delta^{-1}.
\end{equation}

\noindent Taking the linear relationship between \eqref{eq:measurement_model} and $h_{k}$ into account, the full $\MatEng{H}_{k}^{i}$ becomes:

\begin{equation}
    \MatEng{H}_{k}^{i} = \begin{bmatrix}
        \MatEng{H}_{k}^{i, \text{hor}}, 1
    \end{bmatrix}.
\end{equation}

Based on this gradient, each particle is stochastically guided toward the given terrain elevation $\yk$. The numerical calculation of $\MatEng{H}_{k}^{i}$ is also possible, i.e., via finite excitation method. This study, however, advocates the analytic derivation since it requires only one DTED lookup, whereas the numerical one requires at least four. Since the Jacobian needs to be calculated per particle, this saves much computational burden especially when the lookup becomes expensive as the database gets larger or the resolution gets finer.

\subsection{Gaussian Mixture-based Proposal Distribution}\label{sec:GMM_proposal}
By extending the idea of \eqref{eq:random_forcing} to accommodate multiple separate modes from \eqref{eq:measurement_noise}, we propose sampling particles from a mixture proposal distribution \cite{Park2022Evenly}:

\begin{equation} \label{eq:proposal_gmm}
    \begin{aligned}
        \xki 
            &\sim\func{q}{\xk|\x_{k-1}^{i}, \y_{k}} \\
            &= \sum_{n=1}^{N_m} {\pin} 
                \Func{\Gaussian}{\x_{k};
                    \func{f}{\x_{k-1}^{i}} + \MatEng{K}_{k}^{i, (n)}\tilde{\y}_{k}^{i, (n)}, \MatEng{P}_{k|k}^{i, (n)}},
    \end{aligned}
\end{equation}

\noindent where for each mode $n\in{1, 2}$:

\begin{equation} \label{eq:proposal_gmm_gain}
    \begin{aligned}
        \MatEng{K}_{k}^{i, (n)} &= \MatEng{Q}_{k-1}\MatEng{H}_{k}^{i,\transpose} \left(\MatEng{H}_{k}^{i}\MatEng{Q}_{k-1}\MatEng{H}_{k}^{i,\transpose} + \left(\sigman_{k}\right)^{2}\right)^{-1}, \\
        \MatEng{P}_{k|k}^{i, (n)} &= \left(\MatEng{I} - \MatEng{K}_{k}^{i, (n)} \MatEng{H}_{k}^{i}\right)\MatEng{Q}_{k-1}, \\
        \tilde{\y}_{k}^{i, (n)} &= \yk - \left(\hat{\y}_{k}^{i} + \mun_{k}\right).
    \end{aligned}
\end{equation}

\noindent with mixing weights $\pin$ from \eqref{eq:ra_noise}. Here, \eqref{eq:proposal_gmm} should be considered as a \textit{superposition} of a couple of EKF-based random forcing terms, given in \eqref{eq:random_forcing}, where mode selection follows a Bernoulli (binomial) trial with probabilities analogous to those of RA measurement likelihood modes, i.e., $\pi^{(1)}$ and $\pi^{(2)} = 1-\pi^{(1)}$. Each mixture component stochastically aligns with a measurement noise mode from \eqref{eq:measurement_noise} based on the binary outcome.

This design provides more comprehensive support for the posterior distribution. The severe ambiguity in terrain elevation, compounded by multimodal RA characteristics and biases, creates scattered posterior modes. Our mode-specific random forcing approach offers balanced coverage of these modes, unlike simple prior sampling. The method achieves statistical robustness by leveraging the exact RA noise parameters and terrain gradients. Furthermore, the GMM structure generally spreads wider than single Gaussian distributions, providing more stable support for the target distribution.

After a $\xki$ is sampled from \eqref{eq:proposal_gmm}, the weight update follows the exact recursion formula from \eqref{eq:weight_update_general}:

\begin{equation}\label{eq:weight_update_gmm}
    w_{k}^{i}\propto 
        \frac{\func{p_{\VecEng{e}}}{\yk - \func{h}{\xki}}\func{p_{\VecEng{v}}}{\xki - \func{f}{\x_{k-1}^{i}}}w_{k-1}^{i}}{\displaystyle\sum_{n=1}^{N_{m}} {\pin\Func{\mathcal{N}}{\xki;\func{f}{\x_{k-1}^{i}} + \MatEng{K}_{k}^{i, (n)}\tilde{\y}_{k}^{i, (n)}, \MatEng{P}_{k|k}^{i, (n)}}}}.
\end{equation}

\subsection{Augmentation upon Auxiliary Sampling}\label{sec:auxgmm}
We then extend the GMM approach from the previous subsection by incorporating it into the auxiliary variable sampling framework, applying \eqref{eq:proposal_gmm} selectively to particles probable to generate the current measurement $\yk$. Following \eqref{eq:APF_feature}, let us consider jointly sampling both the particle and the index from a proposal density, which satisfies

\begin{equation}
    \func{q}{\cond{\xk, j}{\y_{0:k}}} = \func{q}{\cond{\xk}{j, \y_{0:k}}}\func{q}{\cond{j}{\y_{0:k}}},
\end{equation}

\noindent where the first term is our GMM proposal, i.e., \eqref{eq:proposal_gmm}:

\begin{equation} \label{eq:proposal_design}
    \begin{aligned}
        &\func{q}{\cond{\xk}{j, \y_{0:k}}}= \\ 
        &\qquad\sum_{n=1}^{N_m}{\pin 
        \Func{\Gaussian}{\xk;
            \func{f}{\x_{k-1}^{j}} + \MatEng{K}_{k}^{j, (n)}\tilde{\y}_{k}^{j, (n)}, \MatEng{P}_{k|k}^{j, (n)}}}.
    \end{aligned}
\end{equation}

The proposed sampling method of this study is to combine the above stochastic nudging with the likelihood approximation of $\func{p}{\cond{\yk}{\xk}}$ using $\func{p}{\cond{\yk}{\VecGrk{\varphi}_{k}}}$, just as \eqref{eq:APF_feature}, yielding:

\begin{equation} \label{eq:proposal_auxgmm}
    \begin{aligned}
        &\func{q}{\xk, j|\y_{0:k}} = w_{k-1}^{j}
            \func{p}{\cond{\yk}{\VecGrk{\varphi}_{k}^{j}}}\times\\
        &\qquad  
            \sum_{n=1}^{N_m} {\pin\Func{\Gaussian}{\x_{k};
                \func{f}{\x_{k-1}^{j}} + \MatEng{K}_{k}^{j, (n)}\tilde{\y}_{k}^{j, (n)}, \MatEng{P}_{k|k}^{j, (n)}}},
    \end{aligned}
\end{equation}

\noindent where we use the expectation $\VecGrk{\varphi}_{k}^{i}=\mathbb{E}\left[\xk|\x_{k-1}^{i}\right]$ and discard $j$ as in Section~\ref{sec:APF}.

\begin{figure*}[!htbp]
\hrulefill
\normalsize
\setcounter{equation}{40}
\begin{equation}\label{eq:weight_update_auxgmm}
        w_{k}^{i}
        \propto 
            \frac{
                    \func{p}{\cond{\y_k}{\xki}}
                    \func{p}{\cond{\xki}{\x_{k-1}^{j^i}}}
                }
                {
                    \func{p}{\cond{\y_k}{\VecGrk{\varphi}_{k}^{j^i}}}
                    \displaystyle\sum_{n=1}^{N_{m}} {\pin\Func{\mathcal{N}}{\xki;\func{f}{\x_{k-1}^{j^i}} + \MatEng{K}_{k}^{j^i, (n)}\tilde{\y}_{k}^{j^i, (n)}, \MatEng{P}_{k|k}^{j^i, (n)}}}
                }
\end{equation}
\hrulefill 
\vspace*{4pt} \end{figure*}

The difference of our approach is that, instead of simulating $\xki$ from $\x_{k-1}^{j^i}$ using system motion model \eqref{eq:propagation_model}, the particle is pushed further stochastically towards the measurement using the principle given in Section~\ref{sec:GMM_proposal}. This creates a \textit{nested mixture} structure; outer mixture weights (not particles') follow \eqref{eq:categorical_distribution}, inner mixture weights are $\pin$, and the innermost probability distributions are mode-specific random forcing terms from \eqref{eq:random_forcing}. Then, the weight recursion similar to \eqref{eq:weight_update_gmm} applies using the exact relationship \eqref{eq:weight_update_general}, but the difference is the auxiliary index; see \eqref{eq:weight_update_auxgmm}.

In \eqref{eq:weight_update_auxgmm}, $j^i$ is drawn from a categorical trial whose probability mass function is given by \eqref{eq:categorical_distribution}. Note that in the general APF setup, the last terms of the numerator and denominator cancel out each other in addition to the nullification of the weight of previous time step, $w_{k-1}^{j^i}$, as they are both $\func{p}{\cond{\xki}{\x_{k-1}^{j^i}}}$. 

\begin{figure}[!htbp]
    \centering
    \includegraphics[width=\columnwidth]{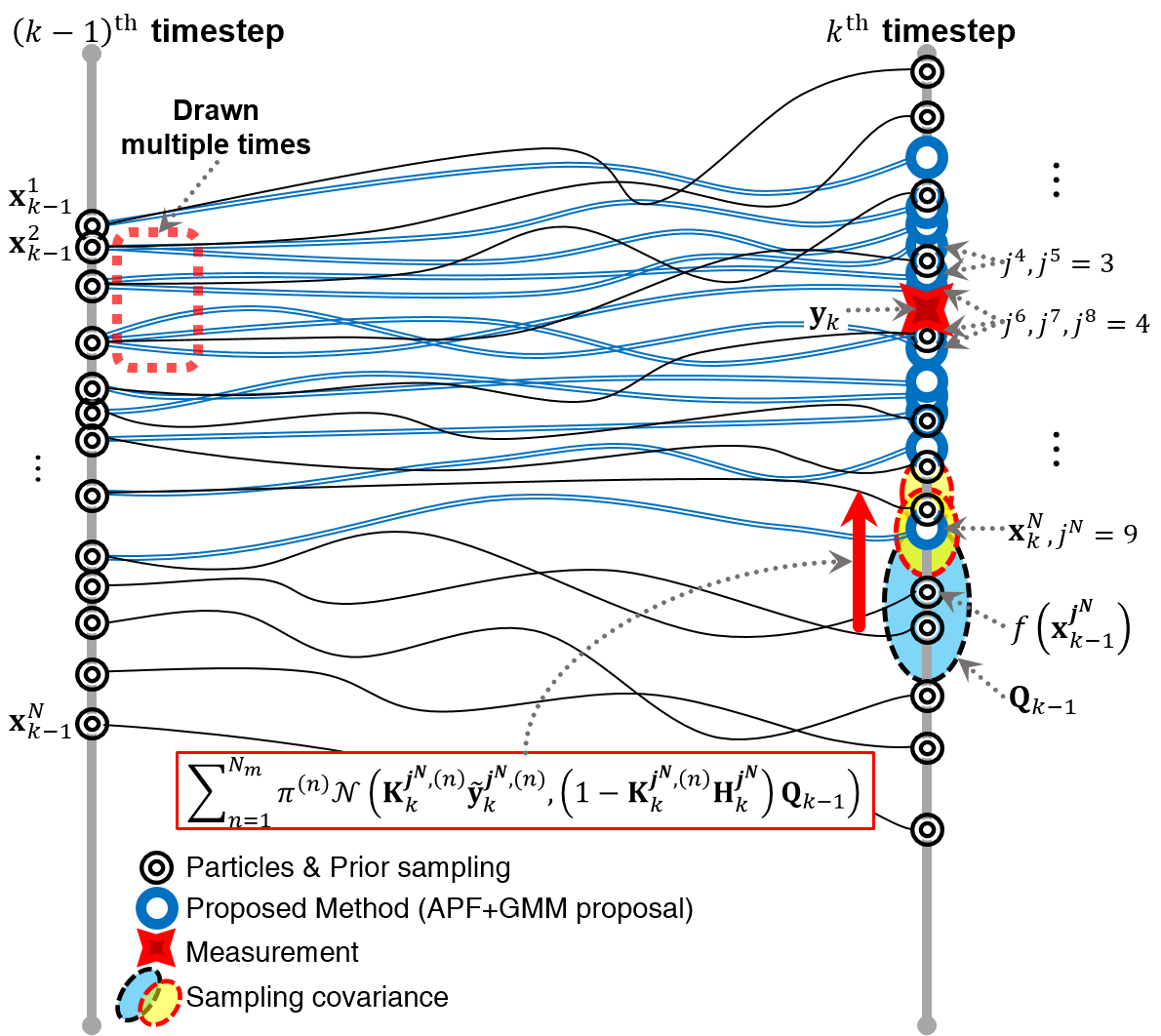}
    \caption{Illustration of sampling process using the proposed proposal distribution.}\label{fig:auxgmm}
\end{figure}

This study aims to push only the probable particles toward the latest available information, while the stochastic nudging has taken the multimodal noise characteristics of $\yk$ into account. 
The proposed sampling method first votes for the probable particles that are more likely to end up around candidate contours, designated by the current RA measurement, and stochastically guides those particles toward the candidate contours. The particles are driven along the direction of terrain gradient, i.e., slope, and the amount of random forcing is determined from \eqref{eq:ekf}. 
Algorithm~\ref{alg:auxgmm} provides the complete implementation details, and Fig.~\ref{fig:auxgmm} highlights the mechanism of the proposed approach. 

In physical sense, the resultant particles gather around at and form \textit{a couple of terrain contours}, each responsible for each modality of RA based on local terrain slope. The auxiliary sampling approach reinforces this effect. We argue that this improves the numerical stability of the derived PF, particularly regarding the weight variance and the effective sample size (ESS) \cite{Arulampalam2002Tutorial, Gustafsson2010Particle}, based on the motivations and rationales given in the previous sections.

\section{Numerical Experiments}\label{sec:experiment}
We evaluate several PF implementations: standard SIR \cite{Arulampalam2002Tutorial}, auxiliary PF (APF) \cite{Pitt1999Filtering}, regularized PF (RPF) \cite{Musso2001Improving}, mixture regularized PF (MRPF) \cite{Murangira2016Mixture}, mode-associated GMM proposal (GMM-A) \cite{Park2022Evenly}, and the proposed augmented GMM (GMM-Aux) using \eqref{eq:proposal_auxgmm}.

In order to validate our claim of improved weight distribution and numerical stability, our analysis focuses on numerical stability indicators. Moreover, the estimator's behavioral trait, such as resilience against unknown time-varying biases, is investigated. These are presented in RMSE form along with the posterior Cram\'er-Rao bound (PCRB) \cite{Tichavsky1998Posterior}. Note, however, that \eqref{eq:measurement_model} is in a non-trivial form and thus is approximated numerically via simultaneous perturbation stochastic approximation (SPSA) \cite{Spall2005Monte} method.

\subsection{Simulation Setup}\label{sec:setup}
Note that $\x^{\text{hor}}$ is curvilinear position, i.e., angles, to comply with a lookup of DTED. In this study, $\xk$ is treated equivalently with its metric counterpart, $\xk^{n}$, based on WGS84 model as

\setcounter{equation}{41}
\begin{equation}\label{eq:llh_to_ned}
    \x^{n} = \MatEng{R}_{e}^{n}\left(\func{\bm{T}_{l}^{e}}{\x}-\func{\bm{T}_{l}^{e}}{\x_{r}}\right),
\end{equation}

\noindent given the reference position $\x_{r}=\x_{0}^{\text{true}}$. The symbols $e$, $n$, and $l$ represent the ECEF, NED, and curvilinear coordinate, respectively, while $\MatEng{R}$ and $\bm{T}$ are coordinate transformations \cite{Groves2013Principlesa}, respectively as
\begin{equation}
    \begin{aligned}
        \MatEng{R}_{e}^{n} &= \begin{bmatrix}
            -S_{L_{r}}C_{\lambda_{r}} & -S_{L_{r}}S_{\lambda_{r}} & C_{L_{r}} \\
            -S_{\lambda_{r}} & C_{\lambda_{r}} & 0 \\
            -C_{L_{r}}C_{\lambda_{r}}& -C_{L_{r}}S_{\lambda_{r}} & -S_{L_{r}}
        \end{bmatrix}, \\
        \bm{T}_{l}^{e}(\x) &= \begin{bmatrix}
            \left(R_E(L) + h\right)C_{L}C_{\lambda} \\
            \left(R_E(L) + h\right)C_{L}S_{\lambda} \\
            \left((1-\epsilon^2)R_E(L) + h\right)S_{L}
    \end{bmatrix},
    \end{aligned}    
\end{equation}
\noindent given the eccentricity of Earth $\epsilon$, the transverse radius of curvature $R_E(L)=a/({1-\epsilon^2\func{\sin^{2}}{L}})^{0.5}$ with $a$ the Earth's equatorial radius.

Given the initial drift $\VecEng{d}^{n}$ represented in NED frame, $\x_0\sim\func{\Gaussian}{\func{\bm{T}_{e}^{l}}{\func{\bm{T}_{l}^{e}}{\x_{r}} + \MatEng{R}_{n}^{e}\VecEng{d}^{n}}, \MatEng{P}_{0}}$, with $\MatEng{P}_{0}$ being the equivalent covariance matrix, i.e., $\func{\text{diag}}{\VecEng{d}^{n}}^2$ in curvilinear coordinates. $\MatEng{Q}_{k}$ can also be treated analogously from its NED equivalent $\MatEng{Q}_{k}^{n}$.

Three measures are used for examining PF instances' internal weight statistics: (1) weight variance, (2) effective sample size (ESS), and (3) Kullback-Leibler divergence \cite{Kullback1951Information} estimator.

The ESS is a widely accepted indicator for particle degeneration, given as
\begin{equation}\label{eq:ESS}
    \text{ESS} = \frac{1}{\sum_{i=1}^{N}{\left(w_k^{i}\right)^2}},
\end{equation}
\noindent and the resampling is triggered adaptively when the ESS is below a certain threshold, e.g., $\text{ESS} < \eta N$, with the design parameter $\eta\in(0, 1]$. It is worth mentioning that $\text{ESS}\in[1, N]$, and the smallest extreme happens when all but one particle has zero weight, while the other end denotes the exact uniformness of the weights, i.e., $1/N$: an equal-weight PF. 

As an alternative indicator for the degeneration, consider the entropy of a discrete probability measure $w_{k}^{1:N}$:
\begin{equation} \label{eq:entropy}
    \Func{H}{w_{k}^{1:N}} = -\sum_{i=1}^{N}{w_{k}^{i}\Func{\log}{w_{k}^{i}}}.
\end{equation}
\noindent At the same time, $\sum_{i=1}^{N}{w^{i}\func{\log}{Nw^{i}}}$ has been proven \cite{Chopin2020Introduction} to be a consistent estimator of
\begin{equation} \label{eq:KLD}
    \begin{aligned}
        D_{\text{KL}}&\left(\func{p}{\cond{\xk}{\yk}}\|\func{q}{\xk}\right)\\
            &\coloneqq\int_{\mathcal{X}_k}{\Func{\log}{\frac{\func{p}{\cond{\xk}{\yk}}}{\func{q}{\xk}}}\func{p}{\cond{\xk}{\yk}}d\xk},
    \end{aligned}
\end{equation}
\noindent which should be the Kullback-Leibler divergence of the proposal distribution relative to the posterior. Therefore, a set of weights with higher entropy, which can be rewritten as
\begin{equation} \label{eq:entropy_equiv}
    \func{H}{w^{1:N}}=\func{\log}{N}-\sum_{i=1}^{N}{w^{i}\func{\log}{Nw^{i}}},
\end{equation}
\noindent should represent a \textit{less divergent} measure from the posterior, given that $\func{\log}{N}$ is a constant. The larger values of the entropy and the ESS indicate less degenerated situations, while the weight variance should get smaller for the equivalent indication.

Then, let $\MatEng{R}_{\text{eq}}$ denote the equivalent variance:
\begin{equation}
    \begin{aligned}
        \MatEng{R}_{\text{eq}} 
            &= \sum_{n}^{N_m} \pin\left((\mun)^2 + (\sigman)^2\right) \\
            &= \pi^{(1)}(\sigma^{(1)})^2 + \pi^{(2)}\left((\mu^{(2)})^2 + (\sigma^{(2)})^2\right),
    \end{aligned}
\end{equation}

\noindent so that $\func{\Gaussian}{\MatEng{0}, \MatEng{R}_{\text{eq}}}$ can approximate \eqref{eq:ra_noise}. We can apply this parameter to interesting analyses, for instance, to the MRPF and to the examination of effectiveness of mode association. In its original version \cite{Murangira2011Robust}, a Gaussian is assumed on $\VecEng{e}_{k}$ during the calculation of MAP per mixture. We can then examine two instances of MRPF: (1) MRPF-1 leveraging negative log likelihood of $\func{\Gaussian}{\VecEng{0}, \MatEng{R}_{\text{eq}}}$ and (2) MRPF-2 using that of \eqref{eq:ra_noise} when calculating MAP via numerical optimization. Note that the Rao-Blackwellization, i.e., marginalization, is not applied here to isolate the effect of sampling, which aligns with the purpose of this study. Thus, they are named MRPF(-MAP), not MRBPF-MAP as used in \cite{Murangira2016Mixture}.

\begin{figure}[!htbp]
    \centering
    \begin{subfigure}{.45\textwidth}
        \includegraphics[width=\textwidth]{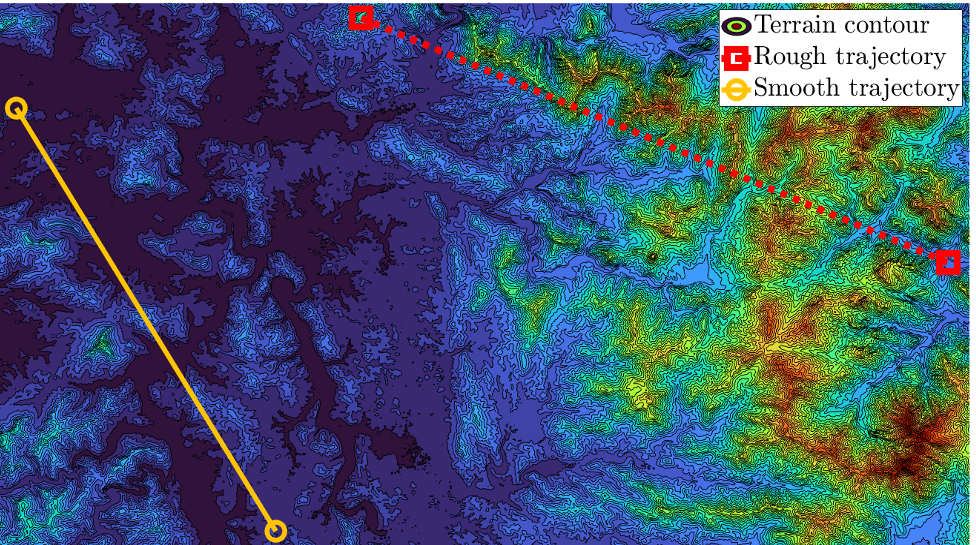}
        \caption{Rough and smooth trajectories.}
        \label{fig:trajectory}
    \end{subfigure}
    \begin{subfigure}{.495\textwidth}
        \includegraphics[width=\textwidth]{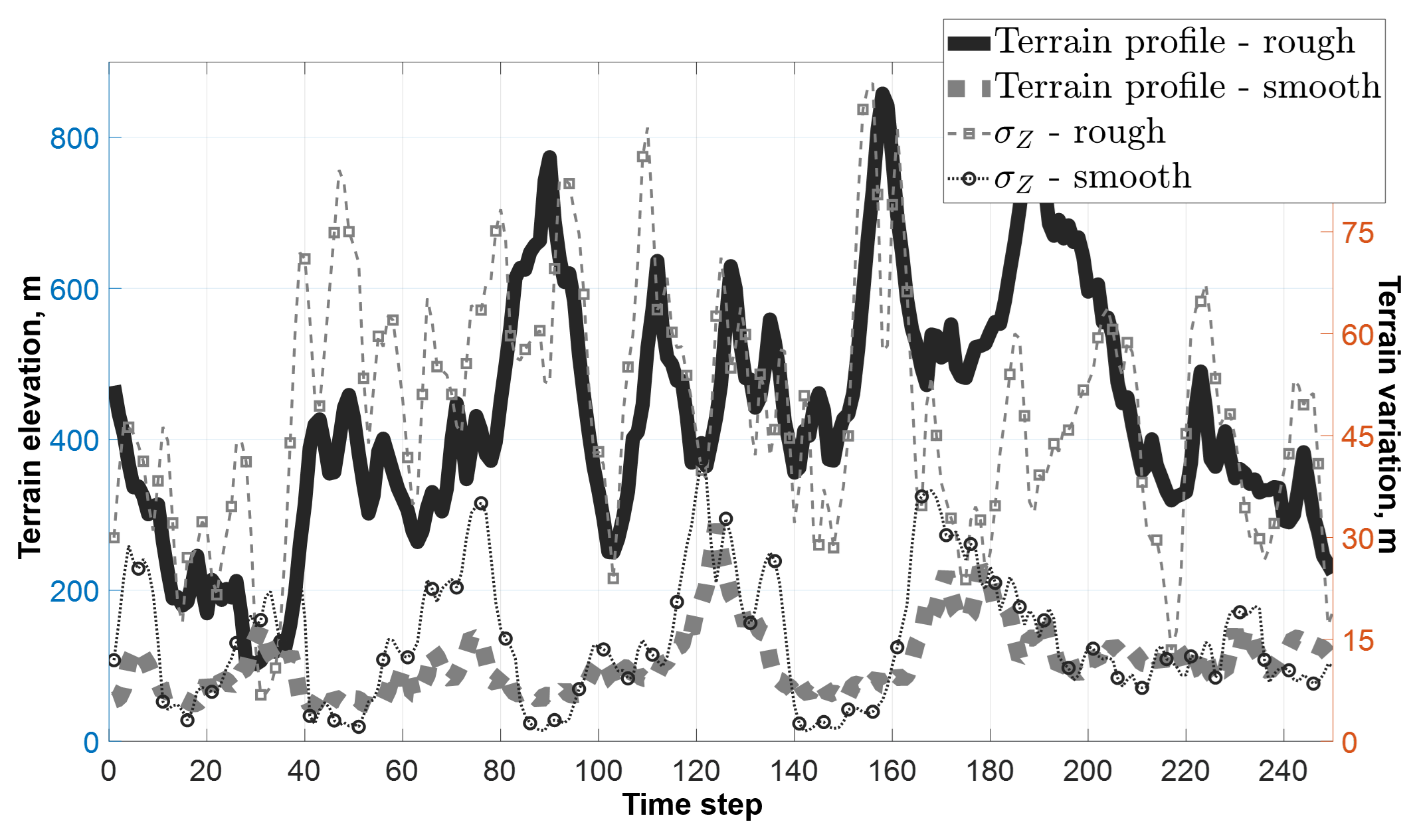}
        \caption{Terrain elevation and roughness profiles along the path.}
        \label{fig:profile}
    \end{subfigure}
    \caption{Simulated trajectories and measurement profiles along trajectories.}
    \label{fig:rough-smooth}
\end{figure}

Two trajectories are tested: rough (R) and smooth (S). They are characterized by roughness $\sigma_{Z}$ \cite{Siouris2004Missile}:
\begin{equation}
    \sigma_{Z} = \sqrt{\frac{1}{N_z-1}\sum_{i=1}^{N_z}{\left(D_i - D\right)^2}},
\end{equation}

\noindent given the number of samples in local area $N_z$, elevation differences $D_i=h_{i}-h_{i+1}$, and the empirical mean $D=\frac{1}{N_z-1}\sum_{i=1}^{N_z-1}{D_i}$. Figure~\ref{fig:rough-smooth} shows these trajectories and elevation profiles.

The acceleration bias profile $s(k)$ (of \eqref{eq:bias_model}) is set to:
\begin{equation}
    s(k) = \begin{cases}
        50, & k\in[50, 75], \\
        105, & k\in[105, 130], \\
        160, & k\in[160, 185], \\
        \infty, & \text{otherwise}.
    \end{cases}
\end{equation}
\noindent Aerial vehicles are often subjected to unknown cyber and/or physical jamming, leading to significant degradation in positioning accuracy. Moreover, IMU faults even deviate the INS incremental whose occurrence is commonly modeled as a step-like function \cite{Avram2017Quadrotor}, which validates the model \eqref{eq:bias_model}. This examines the effect of proper sampling that solidly supports the posterior under deteriorating scenarios. These intervals are highlighted wherever needed. 

A DTED derived from the shuttle radar topography mission (SRTM) \cite{Rabus2003Shuttle} with 1 arcsecond resolution, i.e., $\Delta{L}=\Delta\lambda=1''$ is used, and the remaining simulation parameters are provided in Table~\ref{tab:parameters}. Note that the initial position error of 250m represents the operational range of the proposed sampling approach.

\begin{table}[!htbp]
    \centering
    \caption{Parameters used in the simulation.}
    \label{tab:parameters}
    \begin{tabular}{l c c}
        \toprule
        \textbf{Parameters} & \textbf{Values} & \textbf{Remarks}\\
        \midrule
        $\pi^{(1)}, \pi^{(2)}$ & 0.5, 0.5 & \eqref{eq:measurement_noise} \\
        $\mu^{(1)}, \mu^{(2)}$ & 0, -7.5 & \eqref{eq:measurement_noise}\\
        $\sigma^{(1)}, \sigma^{(2)}$ & 2, 3.5 & \eqref{eq:measurement_noise}\\
        $\MatEng{Q}_{k}^{n}$ & $\func{\text{diag}}{7.5, 7.5, 1.0}^2$ & \eqref{eq:propagation_model} \\
        $\VecEng{b}_{v}^{n}$ & $[1.0, 1.0, 0.2]^{\transpose}$  & \eqref{eq:bias_model}\\
        $\VecEng{b}_{a}^{n}$ & $[0.5, 0.5, 0.1]^{\transpose}$  & \eqref{eq:bias_model} \\
        $\VecEng{d}^{n}$ & $[250\cos{\psi}, 250\sin{\psi}, 25]^{\transpose}$ & $\psi\sim\func{\mathcal{U}}{0, 2\pi}$\\ 
        \bottomrule
    \end{tabular}
\end{table}

\subsection{Simulation Results}\label{sec:result}

\subsubsection{Particle Degeneracy}
First, $M=100$ Monte Carlo simulations are summarized. Particularly, the time averages of the weight variance, entropy of weight, and the ESS are highlighted. The time averages are calculated according to

\begin{equation} \label{eq:time_averaged_ess_h_varw}
    \frac{1}{k_f}\sum_{k=1}^{k_f}{\Func{\overline{\text{var}}}{W_k}}, 
    \frac{1}{k_f}\sum_{k=1}^{k_f}{\Func{\overline{H}}{W_k}},
    \frac{1}{k_f}\sum_{k=1}^{k_f}{\Func{\overline{\text{ESS}}}{W_k}},
\end{equation}

\noindent respectively, where $k_f=250$ is the final time step. Taking the ESS as an example, the barred component is the Monte Carlo mean calculated as
\begin{equation}
    \Func{\overline{\text{ESS}}}{W_{k}} = \frac{1}{M}\sum_{m=1}^{M}{\Func{\text{ESS}^{(m)}}{W_k}}
\end{equation}
\noindent with superscript $(m)$ denoting the result of $m$-th run. The equivalent rule applies to the other two indicators.

Tables \ref{tab:result:variance}, \ref{tab:result:ess}, and \ref{tab:result:entropy} show performance metrics as percentages with respect to their theoretical maxima: $1/N$, $N$ and $\func{\log}{N}$ for weight variance, ESS, and weight entropy, respectively. In each row, the best result is denoted in boldface. Note that lower values are better for weight variance, while higher values indicate more evenly weighted situations for ESS and weight entropy.

\begin{table}[!htbp]
    \centering
    \setlength{\tabcolsep}{3pt}
    \caption{The time average of variance of weights expressed as percentages with respect to the maximum value: $\frac{1}{N}$.}
    \begin{tabular}{c c c c c c c c c c}
        \toprule
        \multirow{2}{*}{Traj.} & \multirow{2}{*}{$\eta$} & \multirow{2}{*}{N} & \multirow{2}{*}{SIR} & \multirow{2}{*}{APF} & \multirow{2}{*}{RPF} & \multicolumn{2}{c}{MRPF} & \multicolumn{2}{c}{GMM} \\
        \cmidrule(l){7-8}\cmidrule(l){9-10} &&&&&& -1& -2& -A& -Aux \\
        \hline
        \multirow{3}{*}{R} & \multirow{3}{*}{0.2}
         & 250 & 24.33 & 12.01 & 24.10 & 5.94 & 26.12 & 22.23 & \textbf{1.18} \\
         && 500 & 18.16 & 9.09 & 19.89 & 4.53 & 20.17 & 18.80 & \textbf{0.69} \\
         && 1000 & 14.13 & 6.78 & 14.18 & 2.98 & 14.26 & 12.59 & \textbf{0.67} \\
        \hline
        \multirow{3}{*}{R} & \multirow{3}{*}{0.7}
         & 250 & 23.81 & 12.98 & 22.02 & 6.12 & 27.53 & 26.80 & \textbf{1.21} \\
         && 500 & 18.73 & 8.88 & 16.15 & 3.82 & 18.65 & 18.81 & \textbf{0.82} \\
         && 1000 & 11.15 & 6.93 & 11.18 & 2.95 & 12.97 & 13.01 & \textbf{0.73} \\
        \hline
        \multirow{3}{*}{S} & \multirow{3}{*}{0.2}
         & 250 & 15.65 & 9.06 & 14.98 & 6.13 & 15.96 & 15.47 & \textbf{0.64} \\
         && 500 & 13.24 & 7.23 & 12.01 & 4.42 & 12.62 & 13.34 & \textbf{0.32} \\
         && 1000 & 9.11 & 4.19 & 9.73 & 3.24 & 10.69 & 9.08 & \textbf{0.23} \\
         \hline
        \multirow{3}{*}{S} & \multirow{3}{*}{0.7}
         & 250 & 14.92 & 9.77 & 14.81 & 5.48 & 14.14 & 14.37 & \textbf{0.69} \\
         && 500 & 10.13 & 6.37 & 10.86 & 3.38 & 12.52 & 12.85 & \textbf{0.40} \\
         && 1000 & 9.30 & 4.46 & 9.14 & 2.44 & 8.98 & 8.99 & \textbf{0.26} \\
        \bottomrule
    \end{tabular}
    \label{tab:result:variance}
\end{table}

\begin{table}[!htbp]
    \centering
    \setlength{\tabcolsep}{3pt}
    \caption{The time average of ESS expressed as percentages with respect to the maximum value: $N$.}
    \begin{tabular}{c c c c c c c c c c}
        \toprule
        \multirow{2}{*}{Traj.} & \multirow{2}{*}{$\eta$} & \multirow{2}{*}{N} & \multirow{2}{*}{SIR} & \multirow{2}{*}{APF} & \multirow{2}{*}{RPF} & \multicolumn{2}{c}{MRPF} & \multicolumn{2}{c}{GMM} \\
        \cmidrule(l){7-8}\cmidrule(l){9-10} &&&&&& -1& -2& -A& -Aux \\
        \hline
        \multirow{3}{*}{R} & \multirow{3}{*}{0.2}
         & 250 & 25.77 & 46.71 & 25.46 & 32.31 & 24.85 & 26.37 & \textbf{58.08} \\
         && 500 & 28.32 & 50.79 & 27.52 & 32.30 & 26.31 & 26.25 & \textbf{59.02} \\
         && 1000 & 29.02 & 51.87 & 28.83 & 32.54 & 27.86 & 29.59 & \textbf{58.08}\\
        \hline
        \multirow{3}{*}{R} & \multirow{3}{*}{0.7}
         & 250 & 32.01 & 45.94 & 34.45 & 42.76 & 32.33 & 30.46 & \textbf{58.32} \\
         && 500 & 35.20 & 50.20 & 38.07 & 44.71 & 33.62 & 32.42 & \textbf{57.76} \\
         && 1000 & 41.49 & 52.67 & 41.19 & 45.78 & 38.00 & 38.62 & \textbf{56.77} \\
        \hline
        \multirow{3}{*}{S} & \multirow{3}{*}{0.2}
         & 250 & 30.40 & 56.08 & 30.86 & 34.37 & 30.00 & 30.19 & \textbf{69.41} \\
         && 500 & 31.12 & 57.59 & 32.00 & 33.77 & 31.31 & 30.31 & \textbf{70.70} \\
         && 1000 & 32.71 & 64.66 & 32.56 & 33.97 & 30.97 & 32.75 & \textbf{70.27} \\
         \hline
        \multirow{3}{*}{S} & \multirow{3}{*}{0.7}
         & 250 & 39.71 & 53.76 & 40.00 & 46.62 & 40.36 & 40.51 & \textbf{68.77} \\
         && 500 & 43.93 & 61.11 & 42.70 & 48.40 & 38.97 & 39.11 & \textbf{69.77} \\
         && 1000 & 43.48 & 64.10 & 43.82 & 47.99 & 44.78 & 44.12 & \textbf{70.26} \\
        \bottomrule
    \end{tabular}
    \label{tab:result:ess}
\end{table}

\begin{table}[!htbp]
    \centering
    \setlength{\tabcolsep}{3pt}
    \caption{The time average of entropy of weights expressed as percentages with respect to the maximum value: $\func{\log}{N}$.}
    \begin{tabular}{c c c c c c c c c c}
        \toprule
        \multirow{2}{*}{Traj.} & \multirow{2}{*}{$\eta$} & \multirow{2}{*}{N} & \multirow{2}{*}{SIR} & \multirow{2}{*}{APF} & \multirow{2}{*}{RPF} & \multicolumn{2}{c}{MRPF} & \multicolumn{2}{c}{GMM} \\
        \cmidrule(l){7-8}\cmidrule(l){9-10} &&&&&& -1& -2& -A& -Aux \\
        \hline
        \multirow{3}{*}{R} & \multirow{3}{*}{0.2}
         & 250 & 57.37 & 77.00 & 57.56 & 76.10 & 55.70 & 58.66 & \textbf{91.56} \\
         && 500 & 63.63 & 81.40 & 61.89 & 78.06 & 60.87 & 61.30 & \textbf{92.92} \\
         && 1000 & 67.58 & 84.26 & 67.54 & 80.70 & 66.55 & 68.73 & \textbf{93.34} \\
        \hline
        \multirow{3}{*}{R} & \multirow{3}{*}{0.7}
         & 250 & 59.89 & 75.79 & 62.49 & 79.41 & 56.91 & 56.20 & \textbf{91.64} \\
         && 500 & 65.22 & 81.25 & 68.47 & 82.83 & 64.26 & 63.22 & \textbf{92.58} \\
         && 1000 & 74.12 & 84.22 & 74.04 & 84.78 & 71.13 & 70.51 & \textbf{92.96} \\
        \hline
        \multirow{3}{*}{S} & \multirow{3}{*}{0.2}
         & 250 & 66.60 & 81.08 & 66.92 & 77.04 & 65.60 & 66.08 & \textbf{93.81} \\
         && 500 & 68.75 & 83.19 & 70.16 & 78.84 & 69.61 & 68.06 & \textbf{94.95} \\
         && 1000 & 73.39 & 88.77 & 72.73 & 80.96 & 71.21 & 73.06 & \textbf{95.37} \\
         \hline
        \multirow{3}{*}{S} & \multirow{3}{*}{0.7}
         & 250 & 69.99 & 79.60 & 70.08 & 81.24 & 70.44 & 70.19 & \textbf{93.65} \\
         && 500 & 75.53 & 85.35 & 74.27 & 84.13 & 71.09 & 70.48 & \textbf{94.64} \\
         && 1000 & 76.18 & 88.51 & 76.36 & 85.39 & 77.26 & 76.10 & \textbf{95.40} \\
        \bottomrule
    \end{tabular}
    \label{tab:result:entropy}
\end{table}

The proposed GMM-Aux approach consistently outperforms other methods across all metrics. These results indicate that the proposed method results in more evenly weighted particles, and thus in a more numerically stable filter. This highlights its robustness in TRN with multimodal measurements and biased odometry. Its performance is most pronounced in challenging scenarios (small particle counts and low resampling rates), indicating superior resilience to weight degeneration. One highlight from the results is that the GMM-Aux case shows 0.23\% of weight variance with respect to the maximum possible (worst) value, which should be almost negligible, while the others show around 2-10\% values.

Increasing particle count $N$ (from 250 to 1000) shows better measures across all methods, reflecting better posterior coverage and reduced weight dominance by individual particles. This trend aligns with the theoretical expectations of PF that larger number of particles provide better coverage of the posterior distribution, reducing the likelihood that few particles dominate very high weights and ensuring higher diversity.

More frequent resampling ($\eta=0.7$) generally improves performance metrics compared to $\eta=0.2$, since it mitigates weight collapse by redistributing particles based on weights. Yet, it is well known that excessive resampling can lead to particle impoverishment. The GMM-Aux effectively balances this tradeoff. It maintains robust performance across both rates, exhibiting the \textit{littlest gap} among different resampling thresholds. Although the auxiliary sampling is often viewed as an implicit resampling at the previous timestep \cite{Gustafsson2010Particle}, the proposed method's effectiveness also stems from controlled randomness that perturbs particles toward probable contours, that might otherwise be under-sampled in APF, ensuring better posterior representation especially in non-analytic terrains corrupted by unknown biases.

\begin{figure}[!htbp]
    \centering
    \includegraphics[width=0.6\columnwidth]{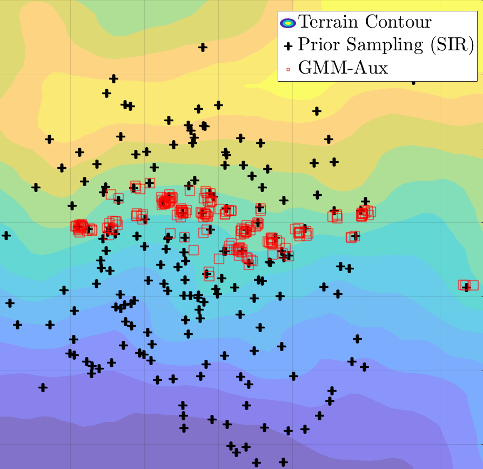}
    \caption{An example of $N=250$ particles sampled by the SIR and GMM-Aux approaches.}
    \label{fig:contraction}
\end{figure}

Meanwhile, high ESS (and small weight variance) does not always guarantee a healthy condition of a PF. For instance, consider that all particles are sampled identically and resultant weights are equally extremely small, which will be normalized to $1/N$ anyway. We claim via Fig.~\ref{fig:contraction} that best indicators for GMM-Aux did not result from particle impoverishment. The figure highlights an example of resultant particles sampled via GMM-Aux (red square) and SIR (black cross) along with terrain elevation map. The particles are sampled along contours while maintaining diversity, evidence of effective modality-seeking behavior. The Gaussian mixture formulation prevents excessive particle contraction through wider and better support of the posterior.

\subsubsection{Influence of Multimodal Parameters}

\begin{figure}[!htbp]
    \centering
    \includegraphics[width=\columnwidth]{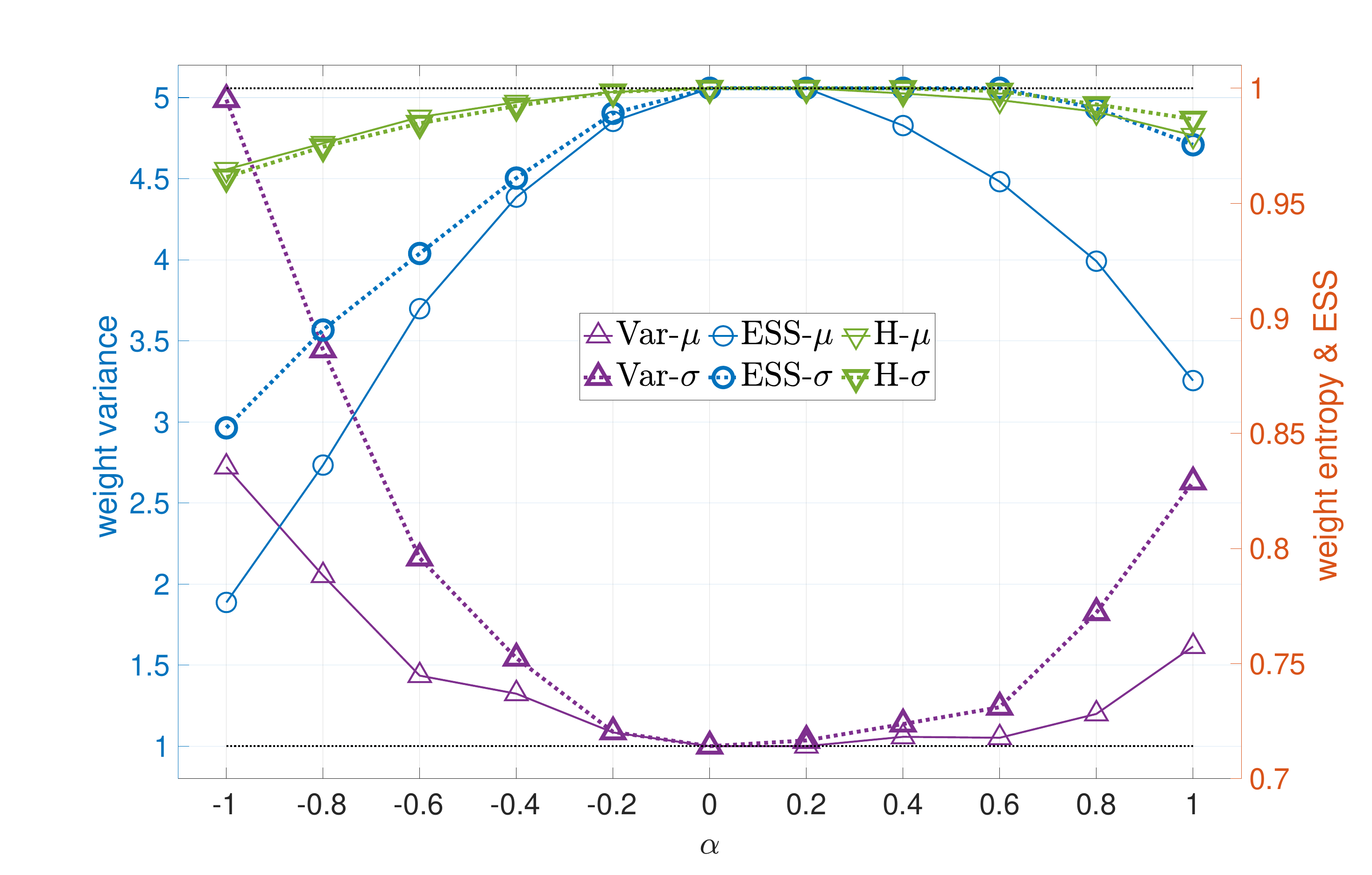}
    \caption{Deviation of PF degeneration measures with respect to the degree of misinformation on the additional mode.}
    \label{fig:parameter_study}
\end{figure}

Then, we introduce a parameter modulation via a factor $\alpha\in[-1, 1]$ such that
\begin{equation}
    \begin{aligned}
        \mu^{(2)'} &= \mu^{(2)} + \alpha\mu^{(2)}, \\
        \sigma^{(2)'} &= 2^{\alpha} \sigma^{(2)}, \\
    \end{aligned}
\end{equation}
\noindent in order to demonstrate the influence of misinformation about the additional mode: $\func{\Gaussian}{\mu^{(2)'}, (\sigma^{(2)'})^2}$. A positive $\alpha$ indicates overestimation of the mode (either more shifted or more varying), while negative values represent the opposite cases. Figure~\ref{fig:parameter_study} summarizes the resultant time-averaged measures under the parameter deviation. The values are relative, i.e., normalized, with respect to the perfect information case: GMM-Aux. 

Generally speaking, as the difference between what the additional mode really is and what the filter knows about it grows, the numerical stability gets degraded. Nevertheless, the effect is admissible when $\alpha$ is between $[-0.2, 0.2]$ yielding nearly 1 to all indicators. When the deviation is larger than 40\%, the deterioration becomes noticeable. For instance, when the variance is identified 40\% smaller ($2^{-0.8}\approx0.6$) than the true value, the average weight variance almost triples and the average ESS shrinks by 10\%. It is also interesting to note that overestimation of the variance of the additional mode (dotted lines, $\alpha>0$) exhibits relatively more stable behavior than the opposite cases, which aligns with our understanding of the PF that too informative likelihood often leads to particle degeneration.

\subsubsection{Resilience Against Unknown Biases}

\begin{figure*}[!htbp]
    \centering
    \includegraphics[width=\linewidth]{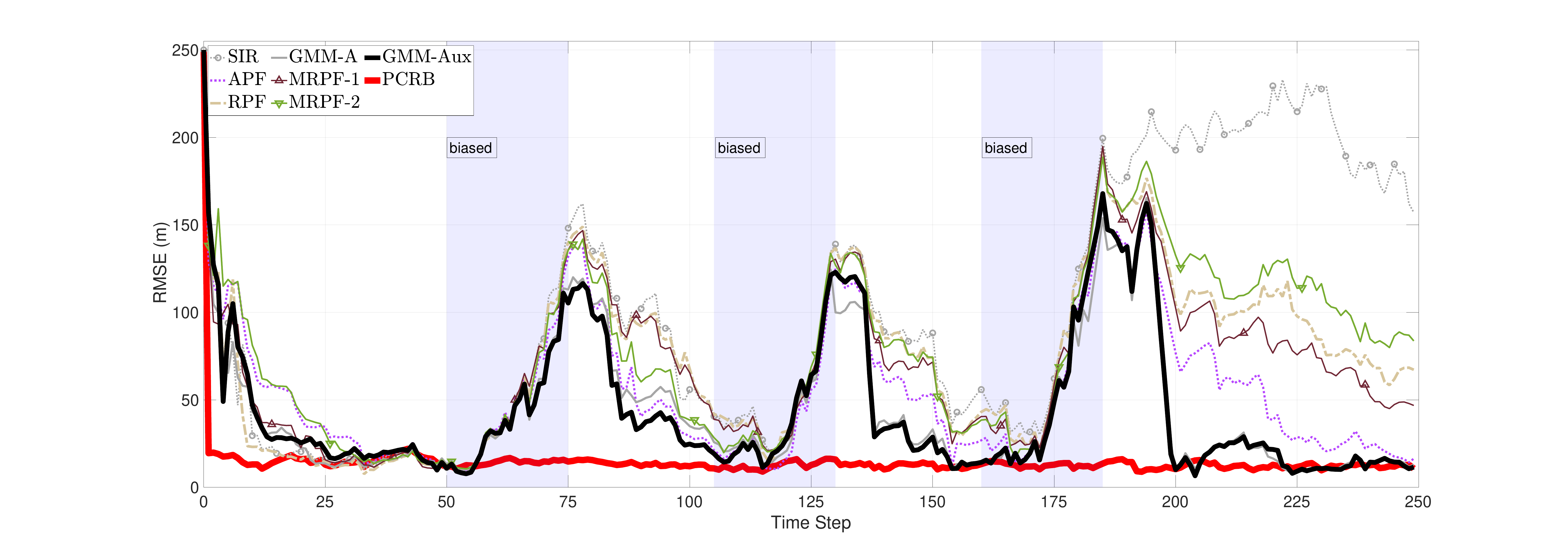}
    \caption{RMSE of horizontal position estimate, comparing multiple PF instances: SIR, APF, RPF, MRPF-1, MRPF-2, GMM-A, and GMM-Aux.}
    \label{fig:horizontal_resilience}
\end{figure*}

\begin{figure*}[!htbp]
    \centering
    \includegraphics[width=\linewidth]{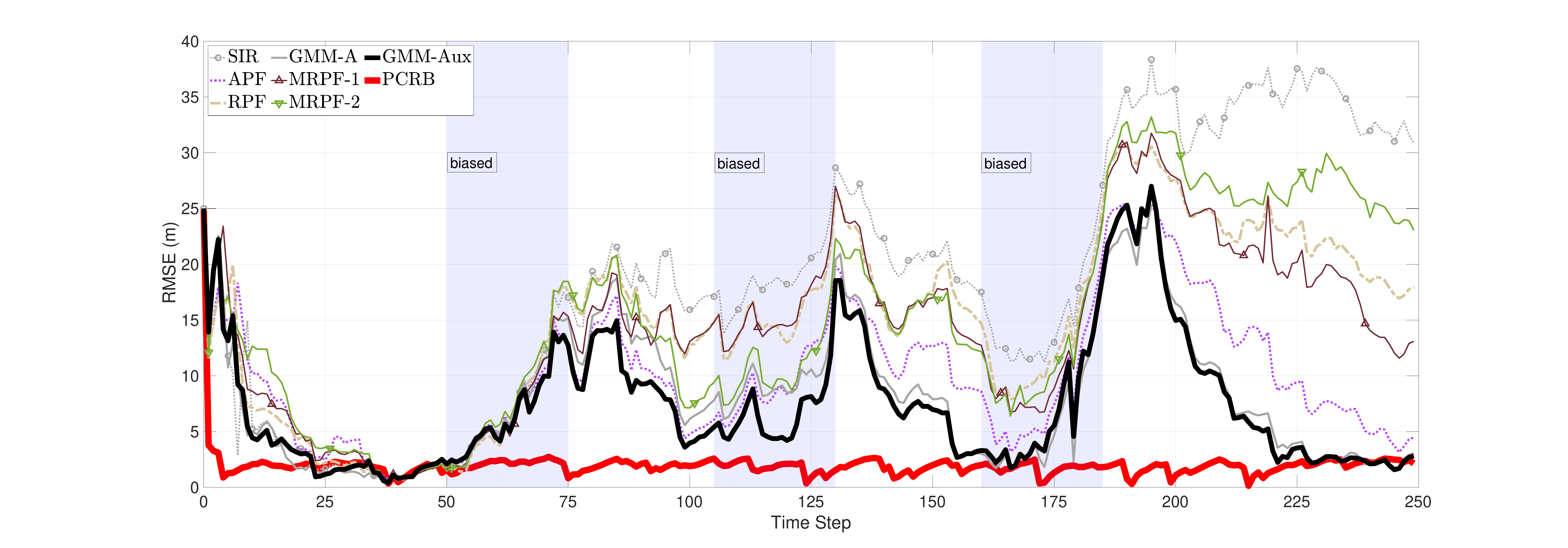}
    \caption{RMSE of vertical position estimate, comparing multiple PF instances: SIR, APF, RPF, MRPF-1, MRPF-2, GMM-A, and GMM-Aux.}
    \label{fig:vertical_resilience}
\end{figure*}

Then, we investigate the estimation accuracy and the derived features of the proposed approach. Particularly, the restoring property of the GMM-Aux approach under the severely deteriorating scenario is reported. Figure~\ref{fig:horizontal_resilience} and \ref{fig:vertical_resilience} present the RMSE of $|\xkhor|$ in NED frame and that of $h_k$ under rough trajectory.

Prior to the occurrence of ramp biases (shaded area, $\VecEng{b}_{a}$), all PF instances converge to the true track, hitting the PCRB. Since $\VecEng{b}_{v}$ is always present in the scenario this study assumes, this highlights stochastically adapting feature of general particle filtering. As $\VecEng{b}_{a}$ accumulates the PF instances gets differentiated, particularly in terms of how they react to the unknown biases. The proposed GMM-Aux approach exhibits the most robust track against this model mismatch. The GMM-A and APF has similar but weaker resiliency, while the regularized PFs (RPF, MRPF) demonstrate even slower recovery. The most plain SIR does not exhibit a converging trend within simulation bound. This phenomenon is observed both on the horizontal plane and the vertical direction.

It is interesting that the recovering feature is strengthened when the multimodality of RA is explicitly considered, either via auxiliary sampling ($\func{p}{\yk|\varphi_k^{j}}$), or explicit mode-associated proposal, i.e., GMM. Note that MRPF-2's sampling is still carried out via single Gaussian with modified covariance, even though the bimodal likelihood is leveraged when calculating MAP.

This swift recovery can be attributed to the method's effective particle redistribution strategy that maintains diversity while tracking probable terrain contours. The intermediate performance of GMM-A without auxiliary sampling suggests that both the mixture model and auxiliary mechanism contribute to the enhanced resilience.  

\begin{figure*}[!htbp]
    \centering
    \includegraphics[width=\linewidth]{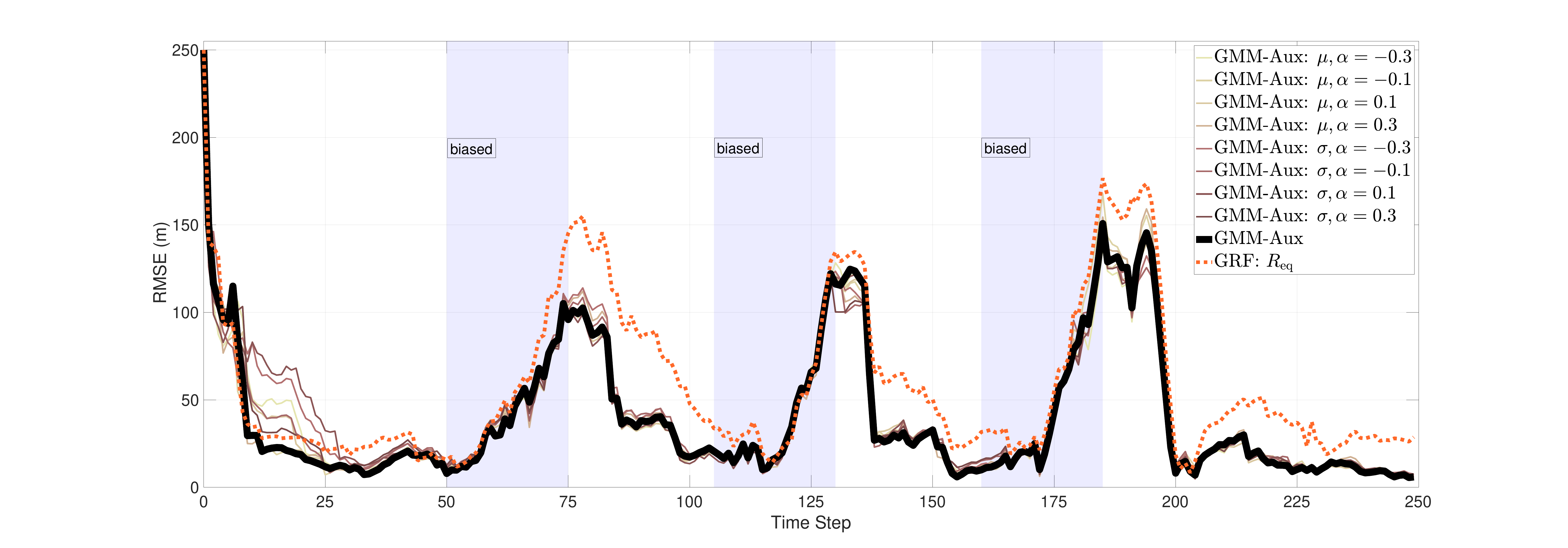}
    \caption{RMSE of horizontal position estimate, comparing GMM-Aux, GMM-Aux parameter deviation, and GRF.}
    \label{fig:horizontal_resilience_grf}
\end{figure*}

\begin{figure*}[!htbp]
    \centering
    \includegraphics[width=\linewidth]{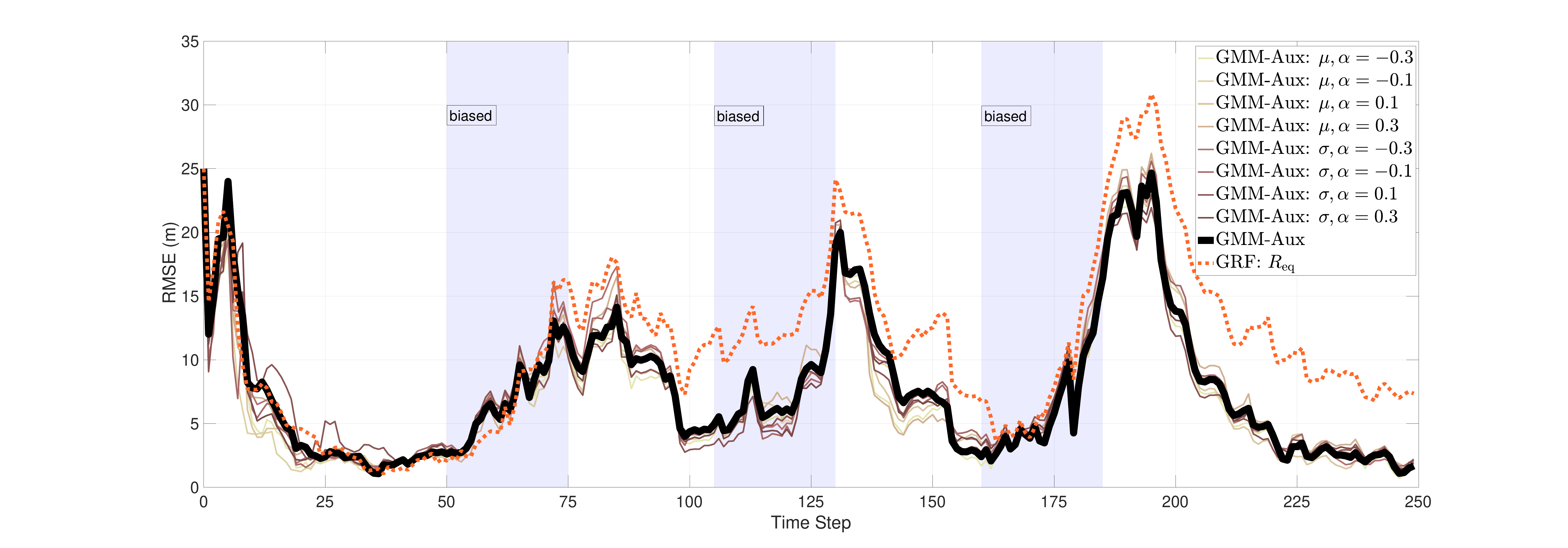}
    \caption{RMSE of vertical position estimate, comparing GMM-Aux, GMM-Aux parameter deviation, and GRF.}
    \label{fig:vertical_resilience_grf}
\end{figure*}

Then, in order to investigate whether this derived feature is driven by the random forcing itself (or not), and to examine the effectiveness of mode association, a set of random forcing families are compared: a Gaussian random forcing (GRF) with $\Func{\Gaussian}{\VecEng{0}, \MatEng{R}_{\text{eq}}}$, GMM-Aux, and GMM-Aux with misinformed parameters. Figure~\ref{fig:horizontal_resilience_grf} and \ref{fig:vertical_resilience_grf} highlight the RMSE of them under the same deteriorating scenario.

All GMM-Aux instances with admissible parameter deviation, i.e., $\alpha\in[-0.3, -0.3]$, exhibit negligible difference in terms of restoring behavior against severe unknown time-varying biases. Only slight deviation on the initial converging course is observed. However, the GRF approach exhibits noticeable lagging of the restoring feature. Although not as severe as SIR nor RPF, it takes longer time to recover the healthy track. The superposition of random forcing \eqref{eq:proposal_auxgmm} allowed for dynamic adaptation to the latest multimodal measurement $\yk$.

\section{Conclusion}\label{sec:conclusion}
This study presented a particle filtering approach for TRN that effectively addressed three critical challenges: multimodal radar altimeter measurements, unknown severe time-varying INS bias, and terrain ambiguity. The proposed method continuously located particles around probable contours, preserving higher effective sample sizes and lower weight variance consistently compared to existing approaches.

Through nested mixture distributions, GMM-Aux improved the effectiveness of individual particles and exhibited notable resilience against unknown time-varying biases through an adaptive forcing mechanism. The demonstrated resilience suggests broader applicability to scenarios with unexpected external disturbances, such as wind gusts, or likelihood with more than two modes. The key insight of exploiting measurement noise structure through mode-associated stochastic forcing, combined with local terrain gradient information, provides a robust framework that could be extended to other navigation problems where measurement ambiguity and state uncertainty pose significant challenges.

While a unimodal forcing was also effective at mitigating biases, fully leveraging multimodal characteristics of radar altimeter measurements is preferable as long as the uncertainty in the modality parameter remains below 30\%. Moreover, The proposed algorithm has been verified for initial position errors of 250 m; batch initialization \cite{Golden1980Terrain} or a conservative filtering---such as the BRPF \cite{Merlinge2019Box}---is recommended before deploying the present method.

Immediate follow-up research should include a mixture operation of GMM-Aux analogous to \cite{Murangira2016Mixture}, as the proposed approach shares a common spirit with the literature's MAP-based importance sampling in drawing particles around measurements. It may be possible to derive an efficient mixture management strategy based on the distinctive traits of the GMM-Aux approach.

\bibliographystyle{IEEEtaes}
\bibliography{TAES2024}

@article{Hong2021Particle,
  title = {Particle {{Filter Approach}} to {{Vision-Based Navigation}} with {{Aerial Image Segmentation}}},
  author = {Hong, Kyungwoo and Kim, Sungjoong and Park, Junwoo and Bang, Hyochoong},
  year = {2021},
  journal = {Journal of Aerospace Information Systems},
  volume = {18},
  number = {12},
  pages = {964--972},
  publisher = {{American Institute of Aeronautics and Astronautics}},
  issn = {2327-3097}
}

@inproceedings{Murangira2011Robust,
  title = {Robust Regularized Particle Filter for Terrain Navigation},
  booktitle = {14th {{International Conference}} on {{Information Fusion}}},
  author = {Murangira, Achille and Musso, Christian and Dahia, Karim and Allard, Jean-Michel},
  year = {2011},
  pages = {1--8},
  publisher = {IEEE},
  isbn = {0-9824438-2-X}
}

@article{Murangira2016Mixture,
  title = {A Mixture Regularized Rao-Blackwellized Particle Filter for Terrain Positioning},
  author = {Murangira, Achille and Musso, Christian and Dahia, Karim},
  year = {2016},
  month = aug,
  journal = {IEEE Transactions on Aerospace and Electronic Systems},
  volume = {52},
  number = {4},
  pages = {1967--1985},
  issn = {0018-9251},
  doi = {10.1109/TAES.2016.150089},
  urldate = {2023-05-30},
  langid = {english}
}

@book{Siouris2004Missile,
  title = {Missile Guidance and Control Systems},
  author = {Siouris, George M.},
  year = {2004},
  publisher = {Springer},
  address = {New York},
  isbn = {978-0-387-00726-7},
  langid = {english},
  lccn = {TL589.4 .S5144 2004}
}

@article{Bergman1999Terrain,
  title = {Terrain Navigation Using {{Bayesian}} Statistics},
  author = {Bergman, Niclas and Ljung, Lennart and Gustafsson, Fredrik},
  year = {1999},
  journal = {IEEE Control Systems Magazine},
  volume = {19},
  number = {3},
  pages = {33--40},
  publisher = {IEEE},
  issn = {1066-033X}
}

@article{Kim2019Approach,
  title = {Approach to Geomagnetic Matching for Navigation Based on a Convolutional Neural Network and Normalised Cross-Correlation},
  author = {Kim, Donghun and Bang, Hyochoong and Lee, Jae Cheul},
  year = {2019},
  journal = {IET Radar, Sonar \& Navigation},
  volume = {13},
  number = {8},
  pages = {1323--1332},
  issn = {1751-8792},
  doi = {10.1049/iet-rsn.2018.5422},
  urldate = {2023-07-27},
  langid = {english}
}

@inproceedings{Park2022Evenly,
  title = {Evenly {{Weighted Particle Filter}} for {{Terrain-referenced Navigation}} Using {{Gaussian Mixture Proposal Distribution}}},
  booktitle = {2022 {{International Conference}} on {{Unmanned Aircraft Systems}} ({{ICUAS}})},
  author = {Park, Junwoo and Bang, Hyochoong},
  year = {2022},
  month = jun,
  pages = {177--183},
  issn = {2575-7296},
  doi = {10.1109/ICUAS54217.2022.9836197}
}

@inproceedings{Park2017New,
  title = {A New Measurement Model of Interferometric Radar Altimeter for Terrain Referenced Navigation Using Particle Filter},
  booktitle = {2017 {{European Navigation Conference}} ({{ENC}})},
  author = {Park, Junwoo and Kim, Youngjoo and Bang, Hyochoong},
  year = {2017},
  month = may,
  pages = {57--64},
  doi = {10.1109/EURONAV.2017.7954173}
}

@article{Doucet2000Sequential,
  title = {On Sequential {{Monte Carlo}} Sampling Methods for {{Bayesian}} Filtering},
  author = {Doucet, Arnaud and Godsill, Simon and Andrieu, Christophe},
  year = {2000},
  journal = {Statistics and computing},
  volume = {10},
  pages = {197--208},
  publisher = {Springer},
  issn = {0960-3174}
}

@book{Chopin2020Introduction,
  title = {An {{Introduction}} to {{Sequential Monte Carlo}}},
  author = {Chopin, Nicolas and Papaspiliopoulos, Omiros},
  year = {2020},
  series = {Springer {{Series}} in {{Statistics}}},
  publisher = {Springer International Publishing},
  address = {Cham},
  doi = {10.1007/978-3-030-47845-2},
  urldate = {2023-05-30},
  isbn = {978-3-030-47844-5 978-3-030-47845-2},
  langid = {english}
}

@book{Doucet2001Sequential,
  title = {Sequential {{Monte Carlo Methods}} in {{Practice}}},
  editor = {Doucet, Arnaud and Freitas, Nando and Gordon, Neil},
  year = {2001},
  publisher = {Springer New York},
  address = {New York, NY},
  doi = {10.1007/978-1-4757-3437-9},
  urldate = {2023-05-30},
  isbn = {978-1-4419-2887-0 978-1-4757-3437-9},
  langid = {english}
}

@article{Musso2001Improving,
  title = {Improving Regularised Particle Filters},
  author = {Musso, Christian and Oudjane, Nadia and Le Gland, Francois},
  year = {2001},
  journal = {Sequential Monte Carlo methods in practice},
  pages = {247--271},
  publisher = {Springer},
  issn = {1441928871}
}

@article{VanLeeuwen2019Particle,
  title = {Particle Filters for High-dimensional Geoscience Applications: {{A}} Review},
  author = {Van Leeuwen, Peter Jan and K{\"u}nsch, Hans R and Nerger, Lars and Potthast, Roland and Reich, Sebastian},
  year = {2019},
  journal = {Quarterly Journal of the Royal Meteorological Society},
  volume = {145},
  number = {723},
  pages = {2335--2365},
  publisher = {Wiley Online Library},
  issn = {0035-9009}
}

@article{Pitt1999Filtering,
  title = {Filtering via {{Simulation}}: {{Auxiliary Particle Filters}}},
  shorttitle = {Filtering via {{Simulation}}},
  author = {Pitt, Michael K. and Shephard, Neil},
  year = {1999},
  month = jun,
  journal = {Journal of the American Statistical Association},
  volume = {94},
  number = {446},
  pages = {590--599},
  issn = {0162-1459, 1537-274X},
  doi = {10.1080/01621459.1999.10474153},
  urldate = {2023-05-30},
  langid = {english}
}

@article{Snyder2015Performance,
  title = {Performance {{Bounds}} for {{Particle Filters Using}} the {{Optimal Proposal}}},
  author = {Snyder, Chris and Bengtsson, Thomas and Morzfeld, Mathias},
  year = {2015},
  month = nov,
  journal = {Monthly Weather Review},
  volume = {143},
  number = {11},
  pages = {4750--4761},
  issn = {0027-0644, 1520-0493},
  doi = {10.1175/MWR-D-15-0144.1},
  urldate = {2023-05-30},
  langid = {english}
}

@article{Li2015Resampling,
  title = {Resampling {{Methods}} for {{Particle Filtering}}: {{Classification}}, Implementation, and Strategies},
  shorttitle = {Resampling {{Methods}} for {{Particle Filtering}}},
  author = {Li, Tiancheng and Bolic, Miodrag and Djuric, Petar M.},
  year = {2015},
  month = may,
  journal = {IEEE Signal Processing Magazine},
  volume = {32},
  number = {3},
  pages = {70--86},
  issn = {1053-5888, 1558-0792},
  doi = {10.1109/MSP.2014.2330626},
  urldate = {2023-05-30},
  langid = {english}
}

@article{Merlinge2019Box,
  title = {A Box Regularized Particle Filter for State Estimation with Severely Ambiguous and Non-Linear Measurements},
  author = {Merlinge, Nicolas and Dahia, Karim and {Piet-Lahanier}, H{\'e}l{\`e}ne and Brusey, James and Horri, Nadjim},
  year = {2019},
  journal = {Automatica},
  volume = {104},
  pages = {102--110},
  publisher = {Elsevier},
  issn = {0005-1098}
}

@article{Merlinge2016Box,
  title = {A {{Box Regularized Particle Filter}} for Terrain Navigation with Highly Non-Linear measurements},
  author = {Merlinge, Nicolas and Dahia, Karim and {Piet-Lahanier}, H{\'e}l{\`e}ne},
  year = {2016},
  journal = {IFAC-PapersOnLine},
  volume = {49},
  number = {17},
  pages = {361--366},
  issn = {24058963},
  doi = {10.1016/j.ifacol.2016.09.062},
  urldate = {2023-05-30},
  langid = {english}
}

@article{Teixeira2017Robust,
  title = {Robust Particle Filter Formulations with Application to Terrain-Aided Navigation},
  shorttitle = {Robust Particle Filter Formulations with Application to Terrain-Aided Navigation},
  author = {Teixeira, Francisco Curado and Quintas, Jo{\~a}o and Maurya, Pramod and Pascoal, Ant{\'o}nio},
  year = {2017},
  month = apr,
  journal = {International Journal of Adaptive Control and Signal Processing},
  volume = {31},
  number = {4},
  pages = {608--651},
  issn = {08906327},
  doi = {10.1002/acs.2692},
  urldate = {2023-05-30},
  langid = {english}
}

@article{Teixeira2012Novel,
  title = {A {{Novel Particle Filter Formulation}} with {{Application}} to {{Terrain-Aided Navigation}}},
  author = {Teixeira, Francisco Curado and Pascoal, Ant{\'o}nio and Maurya, Pramod},
  year = {2012},
  journal = {IFAC Proceedings Volumes},
  volume = {45},
  number = {5},
  pages = {132--139},
  issn = {14746670},
  doi = {10.3182/20120410-3-PT-4028.00023},
  urldate = {2023-05-30},
  langid = {english}
}

@inproceedings{Musso2000Recent,
  title = {Recent Particle Filter Applied to Terrain Navigation},
  booktitle = {Proceedings of the {{Third International Conference}} on {{Information Fusion}}},
  author = {Musso, C. and Oudjane, N.},
  year = {2000},
  pages = {26--33},
  publisher = {IEEE},
  address = {Paris, France},
  doi = {10.1109/IFIC.2000.859835},
  urldate = {2023-05-30},
  isbn = {978-2-7257-0000-7},
  langid = {english}
}

@article{Gustafsson2010Particle,
  title = {Particle Filter Theory and Practice with Positioning Applications},
  author = {Gustafsson, Fredrik},
  year = {2010},
  month = jul,
  journal = {IEEE Aerospace and Electronic Systems Magazine},
  volume = {25},
  number = {7},
  pages = {53--82},
  issn = {0885-8985, 1557-959X},
  doi = {10.1109/MAES.2010.5546308},
  urldate = {2023-05-30},
  langid = {english}
}

@article{Setianto2015Comparison,
  title = {Comparison of {{Kriging}} and {{Inverse Distance Weighted}} ({{IDW}}) {{Interpolation Methods}} in {{Lineament Extraction}} and {{Analysis}}},
  author = {Setianto, Agung and Triandini, Tamia},
  year = {2015},
  month = sep,
  journal = {Journal of Applied Geology},
  volume = {5},
  number = {1},
  issn = {2502-2822},
  doi = {10.22146/jag.7204},
  urldate = {2023-11-20},
  copyright = {Copyright (c) 2022 Agung Setianto, Tamia Triandini},
  langid = {english}
}

@phdthesis{Campbell2006Application,
  title = {Application of Airborne Laser Scanner - Aerial Navigation},
  author = {Campbell, Jacob L.},
  year = {2006},
  address = {United States -- Ohio},
  urldate = {2023-11-20},
  copyright = {Database copyright ProQuest LLC; ProQuest does not claim copyright in the individual underlying works.},
  isbn = {9780542778285},
  langid = {english},
  school = {Ohio University}
}

@book{Groves2013Principlesa,
  title = {Principles of {{GNSS}}, {{Inertial}}, and {{Multi-sensor Integrated Navigation Systems}} (2nd)},
  author = {Groves, Paul},
  year = {2013},
  publisher = {Artech House}
}

@article{Prevot1993Estimating,
  title = {Estimating Surface Soil Moisture and Leaf Area Index of a Wheat Canopy Using a Dual-Frequency ({{C}} and {{X}} Bands) Scatterometer},
  author = {Pr{\'e}vot, Laurent and Champion, I. and Guyot, G.},
  year = {1993},
  month = dec,
  journal = {Remote Sensing of Environment},
  volume = {46},
  number = {3},
  pages = {331--339},
  issn = {0034-4257},
  doi = {10.1016/0034-4257(93)90053-Z},
  urldate = {2023-11-20}
}

@article{Gordon1993Novel,
  title = {Novel Approach to Nonlinear/Non-{{Gaussian Bayesian}} State Estimation},
  author = {Gordon, Neil and Salmond, D.J. and Smith, A.F.M.},
  year = {1993},
  journal = {IEE Proceedings F Radar and Signal Processing},
  volume = {140},
  number = {2},
  pages = {107},
  issn = {0956375X},
  doi = {10.1049/ip-f-2.1993.0015},
  urldate = {2023-05-30},
  langid = {english}
}

@article{Arulampalam2002Tutorial,
  title = {A Tutorial on Particle Filters for Online Nonlinear/Non-{{Gaussian Bayesian}} Tracking},
  author = {Arulampalam, M. Sanjeev and Maskell, Simon and Gordon, Neil and Clapp, Tim},
  year = {Feb./2002},
  journal = {IEEE Transactions on Signal Processing},
  volume = {50},
  number = {2},
  pages = {174--188},
  issn = {1053587X},
  doi = {10.1109/78.978374},
  urldate = {2023-05-30},
  langid = {english}
}

@article{Nordlund2009Marginalized,
  title = {Marginalized {{Particle Filter}} for {{Accurate}} and {{Reliable Terrain-Aided Navigation}}},
  author = {Nordlund, Per-Johan and Gustafsson, Fredrik},
  year = {2009},
  month = oct,
  journal = {IEEE Transactions on Aerospace and Electronic Systems},
  volume = {45},
  number = {4},
  pages = {1385--1399},
  issn = {0018-9251},
  doi = {10.1109/TAES.2009.5310306},
  urldate = {2023-05-30},
  langid = {english}
}

@article{Schon2005Marginalized,
  title = {Marginalized Particle Filters for Mixed Linear/Nonlinear State-Space Models},
  author = {Schon, Thomas B. and Gustafsson, Fredrik and Nordlund, Per-Johan},
  year = {2005},
  month = jul,
  journal = {IEEE Transactions on Signal Processing},
  volume = {53},
  number = {7},
  pages = {2279--2289},
  issn = {1053-587X},
  doi = {10.1109/TSP.2005.849151},
  urldate = {2023-05-30},
  langid = {english}
}

@article{Kullback1951Information,
  title = {On {{Information}} and {{Sufficiency}}},
  author = {Kullback, S. and Leibler, R. A.},
  year = {1951},
  month = mar,
  journal = {The Annals of Mathematical Statistics},
  volume = {22},
  number = {1},
  pages = {79--86},
  issn = {0003-4851},
  doi = {10.1214/aoms/1177729694},
  urldate = {2023-11-20},
  langid = {english}
}

@article{Ly2017Tutorial,
  title = {A {{Tutorial}} on {{Fisher}} Information},
  author = {Ly, Alexander and Marsman, Maarten and Verhagen, Josine and Grasman, Raoul P. P. P. and Wagenmakers, Eric-Jan},
  year = {2017},
  month = oct,
  journal = {Journal of Mathematical Psychology},
  volume = {80},
  pages = {40--55},
  issn = {0022-2496},
  doi = {10.1016/j.jmp.2017.05.006},
  urldate = {2023-11-20}
}

@article{Rabus2003Shuttle,
  title = {The Shuttle Radar Topography Mission---a New Class of Digital Elevation Models Acquired by Spaceborne Radar},
  author = {Rabus, Bernhard and Eineder, Michael and Roth, Achim and Bamler, Richard},
  year = {2003},
  month = feb,
  journal = {ISPRS Journal of Photogrammetry and Remote Sensing},
  volume = {57},
  number = {4},
  pages = {241--262},
  issn = {0924-2716},
  doi = {10.1016/S0924-2716(02)00124-7},
  urldate = {2023-11-20}
}

@article{Jang2017Acquisition,
  title = {Acquisition Method for Terrain Referenced Navigation Using Terrain Contour Intersections},
  author = {Jang, Sukwon and Bang, Hyochoong},
  year = {2017},
  month = sep,
  journal = {IET Radar, Sonar \& Navigation},
  volume = {11},
  number = {9},
  pages = {1444--1450},
  issn = {1751-8792, 1751-8792},
  doi = {10.1049/iet-rsn.2017.0023},
  urldate = {2023-11-20},
  langid = {english}
}

@article{Rosen1996Surface,
  title = {Surface Deformation and Coherence Measurements of {{Kilauea Volcano}}, {{Hawaii}}, from {{SIR}}-{{C}} Radar Interferometry},
  author = {Rosen, Paul A and Hensley, Scott and Zebker, Howard A and Webb, Frank H and Fielding, Eric J},
  year = {1996},
  journal = {Journal of Geophysical Research: Planets},
  volume = {101},
  number = {E10},
  pages = {23109--23125},
  publisher = {Wiley Online Library},
  issn = {0148-0227}
}

@article{Park2020Parameter,
  title = {Parameter {{Estimation}} of {{Radar Noise Model}} for {{Terrain Referenced Navigation Using}} a {{New EM Initialization Method}}},
  author = {Park, Jungmin and Park, Yong-gonjong and Park, Chan Gook},
  year = {2020},
  month = feb,
  journal = {IEEE Transactions on Aerospace and Electronic Systems},
  volume = {56},
  number = {1},
  pages = {107--112},
  issn = {0018-9251, 1557-9603, 2371-9877},
  doi = {10.1109/TAES.2019.2911766},
  urldate = {2023-11-20},
  langid = {english}
}

@article{Bergman1997Bayesian,
  title = {A {{Bayesian Approach}} to {{Terrain-Aided Navigation}}},
  author = {Bergman, Niclas},
  year = {1997},
  month = jul,
  journal = {IFAC Proceedings Volumes},
  volume = {30},
  number = {11},
  pages = {1457--1462},
  issn = {14746670},
  doi = {10.1016/S1474-6670(17)43048-5},
  urldate = {2023-11-20},
  langid = {english}
}

@article{Lee2015Performance,
  title = {Performance Evaluation and Requirements Assessment for Gravity Gradient Referenced Navigation},
  author = {Lee, Jisun and Kwon, Jay Hyoun and Yu, Myeongjong},
  year = {2015},
  journal = {Sensors},
  volume = {15},
  number = {7},
  pages = {16833--16847},
  publisher = {MDPI},
  issn = {1424-8220}
}

@article{Vila-Valls2020Survey,
  title = {Survey on Signal Processing for {{GNSS}} under Ionospheric Scintillation: {{Detection}}, Monitoring, and Mitigation},
  author = {{Vil{\`a}-Valls}, Jordi and Linty, Nicola and Closas, Pau and Dovis, Fabio and Curran, James T},
  year = {2020},
  journal = {NAVIGATION: Journal of the Institute of Navigation},
  volume = {67},
  number = {3},
  pages = {511--535},
  publisher = {Institute of Navigation},
  issn = {0028-1522}
}

@inproceedings{Borio2017Robust,
  title = {Robust Signal Processing for {{GNSS}}},
  booktitle = {2017 {{European Navigation Conference}} ({{ENC}})},
  author = {Borio, Daniele},
  year = {2017},
  pages = {150--158},
  publisher = {IEEE},
  isbn = {1-5090-5922-9}
}

@inproceedings{Koch2006Visionbased,
  title = {A Vision-Based Navigation Algorithm for a {{VTOL-UAV}}},
  booktitle = {{{AIAA Guidance}}, {{Navigation}}, and {{Control Conference}} and {{Exhibit}}},
  author = {Koch, Andreas and Wittig, Hauke and Thielecke, Frank},
  year = {2006},
  pages = {6546}
}

@techreport{Hollowell1990Heli,
  title = {Heli/{{SITAN}}: {{A}} Terrain Referenced Navigation Algorithm for Helicopters},
  author = {Hollowell, Jeff},
  year = {1990},
  institution = {Sandia National Lab.(SNL-NM), Albuquerque, NM (United States)}
}

@inproceedings{Golden1980Terrain,
  title = {Terrain Contour Matching ({{TERCOM}}): A Cruise Missile Guidance Aid},
  booktitle = {Image Processing for Missile Guidance},
  author = {Golden, Joe P},
  year = {1980},
  volume = {238},
  pages = {10--18},
  publisher = {SPIE}
}

@inproceedings{Bergman1997Pointmassa,
  title = {Point-Mass Filter and {{Cramer-Rao}} Bound for Terrain-Aided Navigation},
  booktitle = {Proceedings of the 36th {{IEEE Conference}} on {{Decision}} and {{Control}}},
  author = {Bergman, Niclas and Ljung, Lennart},
  year = {1997},
  volume = {1},
  pages = {565--570},
  publisher = {IEEE},
  isbn = {0-7803-4187-2}
}

@article{Park2024Visual,
  title = {Visual {{Semantic Context}} and {{Efficient Map-Based Rotation-Invariant Estimation}} of {{Position}} and {{Heading}}},
  author = {Park, Junwoo and Kim, Sungjoong and Hong, Kyungwoo and Bang, Hyochoong},
  year = {2024},
  journal = {NAVIGATION: Journal of the Institute of Navigation},
  volume = {71},
  number = {1},
  pages = {navi.634},
  issn = {0028-1522, 2161-4296},
  doi = {10.33012/navi.634},
  urldate = {2024-09-08},
  langid = {english}
}

@article{Tichavsky1998Posterior,
    title = {Posterior {Cramer}-{Rao} bounds for discrete-time nonlinear filtering},
    volume = {46},
    issn = {1053587X},
    doi = {10.1109/78.668800},
    language = {en},
    number = {5},
    urldate = {2023-05-30},
    journal = {IEEE Transactions on Signal Processing},
    author = {Tichavsky, Petr and Muravchik, Carlos and Nehorai, Arye},
    month = may,
    year = {1998},
    pages = {1386--1396},
}

@article{Spall2005Monte,
  title = {Monte {{Carlo Computation}} of the {{Fisher Information Matrix}} in {{Nonstandard Settings}}},
  author = {Spall, James C},
  year = {2005},
  month = dec,
  journal = {Journal of Computational and Graphical Statistics},
  volume = {14},
  number = {4},
  pages = {889--909},
  issn = {1061-8600, 1537-2715},
  doi = {10.1198/106186005X78800},
  urldate = {2023-05-30},
  langid = {english}
}

@article{Avram2017Quadrotor,
  title={Quadrotor sensor fault diagnosis with experimental results},
  author={Avram, Remus C and Zhang, Xiaodong and Muse, Jonathan},
  journal={Journal of Intelligent \& Robotic Systems},
  volume={86},
  pages={115--137},
  year={2017},
  publisher={Springer}
}

@article{Li2017Particle,
  title={Particle filtering with invertible particle flow},
  author={Li, Yunpeng and Coates, Mark},
  journal={IEEE Transactions on Signal Processing},
  volume={65},
  number={15},
  pages={4102--4116},
  year={2017},
  publisher={IEEE}
}

@inproceedings{Musso2019Terrain,
  title={Terrain-aided navigation with an atomic gravimeter},
  author={Musso, Christian and Sacleux, Bernard and Bresson, Alexandre and Allard, Jean-Michel and Dahia, Karim and Bidel, Yannick and Zahzam, Nassim and Palmier, Camille},
  booktitle={2019 22th International Conference on Information Fusion (FUSION)},
  pages={1--8},
  year={2019},
  organization={IEEE}
}

\begin{IEEEbiography}[{\includegraphics[width=1in,height=1.25in,clip,keepaspectratio]{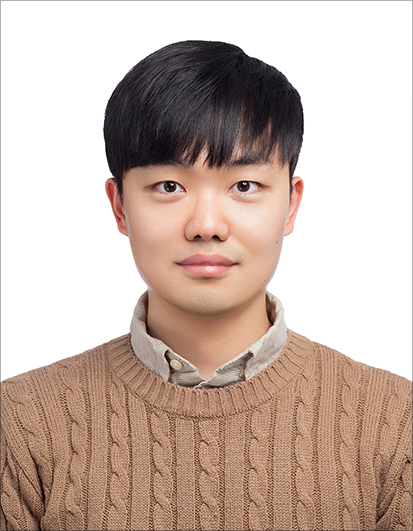}}]{Junwoo Park} received his B.S., M.S., and Ph.D. degrees in aerospace engineering from Korea Advanced Institute of Science and Technology, Daejeon, South Korea, in 2016, 2018, and 2024, respectively. He is currently an aerospace engineer in Nearthlab, Inc., Seoul, South Korea. His research interests include database-referenced navigation, unmanned aerial vehicle navigation, nonlinear state estimation, and particle filtering.
\end{IEEEbiography}

\begin{IEEEbiography}[{\includegraphics[width=1in,height=1.25in,clip,keepaspectratio]{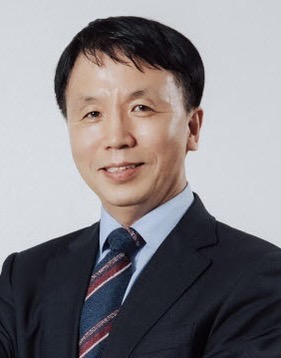}}]{Hyochoong Bang} (Member, IEEE), received the B.S. and M.S. degrees in aeronautical engineering from Seoul National University, Seoul, South Korea, in 1985 and 1987, respectively, and the Ph.D. degree from Texas A\&M University, College Station, in 1992. 

From 1992 to 1994, he was a Research Assistant Professor with the U.S. Naval Postgraduate School, Monterey, CA, where he conducted spacecraft attitude control research. From 1995 to 1999, he was with Korea Aerospace Research Institute, Daejeon, South Korea. Since 2001, he has been a Professor with Korea Advanced Institute of Science and Technology, Daejeon. His current research interests include spacecraft guidance and attitude control, and unmanned aircraft guidance and flight control systems design.
\end{IEEEbiography}

\section*{Appendix}\label{sec:appendix}
\subsection{Pseudo code of the proposed approach (GMM-Aux)}\label{sec:appendix:pseudo-code}

\begin{algorithm*}[!hp]
\caption{Proposed Particle Filter (GMM Proposal with Auxiliary Sampling)}\label{alg:auxgmm}
\begin{algorithmic}
\REQUIRE $\left\{\x_{k-1}^{i}, w_{k-1}^{i}\right\}_{i=1}^{N}, \yk$
\FOR{$i=1:N$}
\STATE{Calculate~~$\VecGrk{\varphi}_{k}^{i}=\mathbb{E}\left[\cond{\xk}{\x_{k-1}^{i}}\right]$ and $\lambda_{k}^{i}=\func{p}{\cond{\yk}{\VecGrk{\varphi}_{k}^{i}}}w_{k-1}^{i}$}
\ENDFOR
\STATE{Initialize an index set, $\mathcal{I}\gets\emptyset$, and normalize first-stage weights: $\{\lambda_{k}^{i}\}_{i=1}^{N}$}
\FOR{$i=1:N$}
\STATE{$\lambda_{k}^{i} \gets \linefrac{\lambda_{k}^{i}}{\sum_{i'=1}^{N}{\lambda_{k}^{i'}}}$}
\ENDFOR
\FOR{$i=1:N$}
\STATE{Draw $j^{i}\sim\Func{p}{\mathcal{J}}$, where~$\Func{p}{\mathcal{J}=j}=\lambda_{k}^{j}$}
\STATE{$\mathcal{I}\gets\mathcal{I}\cup\{j^i\}$}
\ENDFOR
\FOR{$j\in\mathcal{I}$}
\STATE{Calculate~~$\MatEng{H}_{k}^{j}= \left.\PD[line]{\Func{h}{\xk}}{\xk}\right|_{\xk=\func{f}{\x_{k-1}^{j}}}$} \COMMENT{recall: \eqref{eq:Jacobian}}
\FOR{$n\in\{1,2\}$}
\STATE{Calculate~~$\MatEng{K}_{k}^{j,(n)} = \MatEng{Q}_{k-1}\MatEng{H}_{k}^{j,\transpose} \left(\MatEng{H}_{k}^{j}\MatEng{Q}_{k-1}\MatEng{H}_{k}^{j,\transpose} + (\sigman_{k})^{2}\right)^{-1}$}
\STATE{Calculate~~$\MatEng{P}_{k|k}^{j, (n)} = \left(\MatEng{I} - \MatEng{K}_{k}^{j, (n)} \MatEng{H}_{k}^{j}\right)\MatEng{Q}_{k-1}$ and $\tilde{\y}_{k}^{j, (n)} = \y_{k} - \hat{\y}_{k}^{j} + \mun_{k}$} \COMMENT{recall: \eqref{eq:proposal_gmm_gain}}
\ENDFOR
\ENDFOR
\FOR{$i=1:N$}
\STATE{Draw $\xki\sim \Func{q}{\cond{\xk}{j^i, \y_{0:k}}} = 
    \sum_{n=1}^{N_m} {\pin} 
        \Func{\mathcal{N}}{\x_{k};
            \func{f}{\x_{k-1}^{j^i}} + \MatEng{K}_{k}^{j^i, (n)}\tilde{\y}_{k}^{j^i, (n)}, \MatEng{P}_{k|k}^{j^i, (n)}}$} \COMMENT{recall: \eqref{eq:proposal_auxgmm}}
\STATE{Assign weight $w_k^{i}=\frac{
                    \func{p_{\VecEng{e}}}{\yk - \func{h}{\xki}}
                    \func{p_{\VecEng{v}}}{\xki - \func{f}{\x_{k-1}^{j^i}}}
                }
                {
                    \func{p_{\VecEng{e}}}{\yk - \func{h}{\VecGrk{\varphi}_{k}^{j^i}}}
                    \sum_{n=1}^{N_{m}} {\pi^{(n)}\Func{\mathcal{N}}{\xki;\func{f}{\x_{k-1}^{j^i}} + \MatEng{K}_{k}^{j^i, (n)}\tilde{\y}_{k}^{j^i, (n)}, \MatEng{P}_{k|k}^{j^i, (n)}}}
                }$} \COMMENT{recall: \eqref{eq:weight_update_auxgmm}}
\ENDFOR
\FOR{$i=1:N$}
\STATE{$w_{k}^{i} \gets \linefrac{w_{k}^{i}}{\sum_{i'=1}^{N}{w_{k}^{i'}}}$}
\ENDFOR
\STATE{\RETURN{$\left\{\xki, w_{k}^{i}\right\}_{i=1}^{N}=\text{Resampling}\left(\left\{\xki, w_{k}^{i}\right\}_{i=1}^{N},\eta\right)$}}
\end{algorithmic}
\end{algorithm*}
\end{document}